\documentclass[journal=jacsat,manuscript=article]{achemso}
\setkeys{acs}{articletitle = true}
\setkeys{acs}{chaptertitle = true}
\setkeys{acs}{maxauthors = 0}
\setkeys{acs}{keywords = true}
\usepackage{chemformula} 
\usepackage[T1]{fontenc} 
\usepackage{braket}
\usepackage{ulem}

\author{Wataru Ota}
\affiliation{Fukui Institute for Fundamental Chemistry, Kyoto University, Sakyo-ku, Kyoto 606-8103, Japan}
\alsoaffiliation{Department of Molecular Engineering, Graduate School of Engineering, Kyoto University, Nishikyo-ku, Kyoto 615-8510, Japan}

\author{Yasuro Kojima}
\affiliation{Department of Molecular Engineering, Graduate School of Engineering, Kyoto University, Nishikyo-ku, Kyoto 615-8510, Japan}

\author{Saburo Hosokawa}
\affiliation{Department of Molecular Engineering, Graduate School of Engineering, Kyoto University, Nishikyo-ku, Kyoto 615-8510, Japan}
\alsoaffiliation{
Elements Strategy Initiative for Catalysts \& Batteries (ESICB), Kyoto University, Kyotodaigaku Katsura, Nishikyo-ku, Kyoto 615-8245, Japan
}

\author{Kentaro Teramura}
\affiliation{Department of Molecular Engineering, Graduate School of Engineering, Kyoto University, Nishikyo-ku, Kyoto 615-8510, Japan}
\alsoaffiliation{
Elements Strategy Initiative for Catalysts \& Batteries (ESICB), Kyoto University, Kyotodaigaku Katsura, Nishikyo-ku, Kyoto 615-8245, Japan
}

\author{Tsunehiro Tanaka}
\affiliation{Department of Molecular Engineering, Graduate School of Engineering, Kyoto University, Nishikyo-ku, Kyoto 615-8510, Japan}
\alsoaffiliation{
Elements Strategy Initiative for Catalysts \& Batteries (ESICB), Kyoto University, Kyotodaigaku Katsura, Nishikyo-ku, Kyoto 615-8245, Japan
}

\author{Tohru Sato}
\affiliation{Fukui Institute for Fundamental Chemistry, Kyoto University, Sakyo-ku, Kyoto 606-8103, Japan}
\alsoaffiliation{Department of Molecular Engineering, Graduate School of Engineering, Kyoto University, Nishikyo-ku, Kyoto 615-8510, Japan}
\alsoaffiliation{
Elements Strategy Initiative for Catalysts \& Batteries (ESICB), Kyoto University, Kyotodaigaku Katsura, Nishikyo-ku, Kyoto 615-8245, Japan
}
\email{tsato@scl.kyoto-u.ac.jp}

\title{
Role of Catalyst Support and
Regioselectivity of Molecular Adsorption on a Metal Oxide Surface:
NO Reduction on Cu/$\gamma$-Alumina
}

\abbreviations{SSHT,IR,UV/Vis,VCC,VCD,OVCC,OVCD,HOMO,LUMO,SOMO}
\keywords{
$\gamma$-alumina, 
NO reduction, 
anchoring effect,
regioselectivity,
vibronic couplings
}

\begin{document}

\begin{tocentry}
	\vspace{-0.13cm}\hspace{0.04cm}\includegraphics[width=1\hsize]{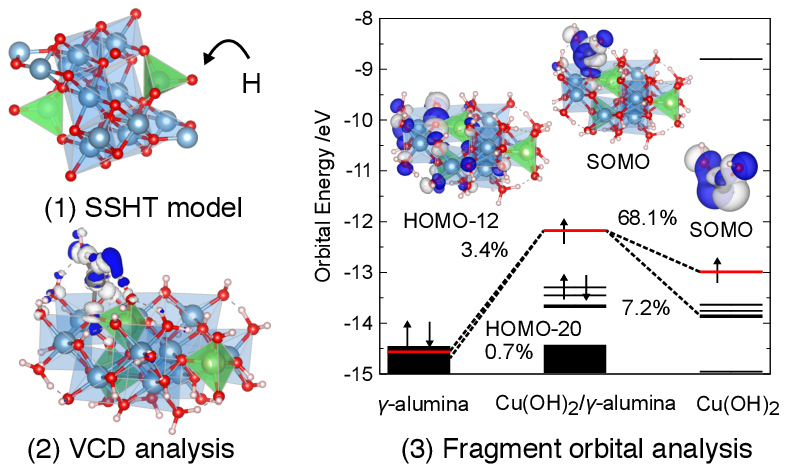}
\end{tocentry}

\begin{abstract}
The role of catalyst support and 
regioselectivity of molecular adsorption on a metal oxide surface
is investigated for the NO reduction on a Cu/$\gamma$-alumina heterogeneous catalyst.
For the solid surface,
computational models of the $\gamma$-alumina surface are constructed 
based on the Step-by-Step Hydrogen Termination (SSHT) approach.
Dangling bonds, 
which appear by cutting the crystal structure of a model,
are terminated stepwise with H atoms
until the model has an appropriate energy gap.
The obtained SSHT models exhibit the realistic infrared (IR) and ultraviolet-visible (UV/Vis) spectra.
Vibronic coupling density (VCD), as a reactivity index,
is employed to elucidate the regioselectivity
of the Cu adsorption on the $\gamma$-alumina
and that of the NO adsorption on the Cu/$\gamma$-alumina
in place of the frontier orbital theory that could not provide clear results.
We discovered that the highly dispersed Cu atoms are
loaded on Lewis-basic O atoms, which is known as {\it anchoring effect},
located in the tetrahedral sites of the $\gamma$-alumina surface.
The role of the $\gamma$-alumina support is 
to raise the frontier orbital of the Cu catalyst,
which in turn gives rise to the electron back-donation from the Cu/$\gamma$-alumina to NO.
In addition,
the penetration of the VCD distribution of the Cu/$\gamma$-alumina 
into the $\gamma$-alumina support indicates
that the excessive reaction energies dissipate into the support 
after the NO adsorption and reduction.
In other words, the support plays the role of a heat bath.
The NO reduction on the Cu/$\gamma$-alumina proceeds even in an oxidative atmosphere
because the Cu--NO bond is strongly bounded 
compared to the Cu--O$_2$ bond.
\end{abstract}

\section{Introduction}

Three-way catalysts are used in automobiles 
to remove nitrogen oxides, carbon monoxide, and hydrocarbons from exhaust gas.
Typical components of the three-way catalysts are 
platinum-group metal (PGM) species such as Rh, Pd, and Pt.
In particular, Rh is responsible for reducing nitrogen oxides to N$_2$
\cite{Kavspar2003_419,Shelef1994_433}.
Thus, the NO adsorption on the Rh surface has been extensively studied
\cite{Brown2000_2578}.
In addition to the experiments,
theoretical calculations have also attempted to reveal 
the NO reduction processes on the Rh catalysts
\cite{Deushi2017_15272,Ishikawa2018_17378,Takagi2019_7021}.
Nevertheless, the reduction mechanism, 
especially the role of the catalyst support, 
remains unknown.
For example,
Ward \textit{et al.} investigated the electronic states of (NO)$_2$ 
on the Rh, Pd, and Pt surfaces using the extended H\"uckel theory
\cite{Ward1993_7691}.
Electron back-donation occurs from the Rh $d$ bands to the (NO)$_2$ $2\pi$ orbital
because the $d$ bands are energetically close to the $2\pi$ orbital.
The $2\pi$ orbital has a bonding character in the N--N bond 
and anti-bonding character in the N--O bonds.
Consequently, 
the N--N bond is strengthened while the N--O bonds are weakened owing to back-donation,
which is suitable for the reduction of NO to N$_2$.
In contrast to the Rh surface, 
the back-donation hardly occurs for the Pd and Pt surfaces
because these $d$ bands are energetically much lower than the (NO)$_2$ $2\pi$ orbital.
Thus, the (NO)$_2$ adsorption on the catalysts can be important
as an initial step of the NO reduction.
Ward \textit{et al}. also reported that
the $\alpha$-alumina support varies the Fermi level of the Rh, Pd, and Pt surfaces,
which affects the strength of N--O bond on their surfaces
\cite{Ward1993_85}.

The replacement for Rh with a ubiquitous element such as Cu is desirable
because Rh is an expensive and rare metal.
Cu/$\gamma$-alumina is a heterogeneous catalyst
that reduces nitrogen oxides with high efficiency
\cite{Yamamoto2002_2449,Yamamoto2002_113,Amano2006_282,Hosokawa2017_74,Fukuda2018_3833}.
The most important feature of Cu catalyst is that
NO reduction proceeds in an oxidative atmosphere,
whereas PGM catalyst hardly exhibits catalytic activity under such a condition.
In other words,
NO can be easily adsorbed on the Cu/$\gamma$-alumina even in the presence of O$_2$.
The active site for the reduction is determined to be the highly dispersed Cu$^{2+}$ species 
\cite{Yamamoto2002_2449}.
The degree of catalyst dispersion generally depends on 
the strength of the interaction between a catalyst and its support,
which is known as an anchoring effect
\cite{Nagai2006_103,Machida2014_5799}.
$\gamma$-alumina,
one of transitional phases of the alumina, 
has been extensively used as a catalyst support
\cite{Trueba2005_3393}.
Owing to the existence of various types of adsorption sites 
on its surfaces due to its amorphous structure,
the adsorption sites of the Cu catalyst on the $\gamma$-alumina as well as
those of the NO on the Cu/$\gamma$-alumina are not clear.
Therefore, 
the regioselectivity of the adsorptions require clarification
to investigate the mechanism of the NO reduction on the Cu$/\gamma$-alumina
while considering the role of the catalyst support.

The identification of reactive sites on a molecule or solid surface
is one of the important problems in quantum chemistry.
The frontier orbital theory evaluates the stabilization arising from the charge-transfer interactions,
and has been successful in clarifying the regioselectivity of chemical reactions
\cite{Fukui1952_722,Fukui1982_747}.
However, this theory sometimes has difficulty 
in predicting reactive sites of a large system such as a solid surface
due to the delocalization of frontier orbitals.
This is because only the effect of the electronic states is considered.
In addition to the stabilization via charge transfer interaction, 
further stabilization arises from structural relaxation caused by vibronic couplings.
Vibronic coupling density (VCD), 
which is calculated from electronic and vibrational states,
identifies reactive sites 
where the stabilization by vibronic couplings is significant in the course of a reaction
\cite{Sato2008_758}.
The VCD analyses have successfully predicted the regioselectivity of 
the CO$_2$ and H$_2$ adsorption on the gallium oxide surface 
because of the localization of the vibrational states
\cite{Kojima2019_239,Ota2018_138}.
In addition, the VCD can be used as a reactivity index 
for the fullerenes
\cite{Sato2012_257,Haruta2012_9702,Haruta2013_012003,Haruta2014_3510,Haruta2014_141}
and aromatic hydrocarbons
\cite{Haruta2015_590},
which are other examples where the frontier orbital theory fails to predict the regioselectivity.
On the basis of the VCD analysis,
the regioselectivity of the molecular adsorption 
has been predicted from the calculations of only the solid surface.
In other words,
there is no need to find an adsorption site with the minimum energy
by calculating the energies of all possible molecular arrangements on a solid surface
at the expense of computational costs.

For the VCD analysis for a solid surface, 
a computational model is required.
Dangling bonds are generated on the metal oxide surfaces
after simply cutting its three-dimensional (3D) crystal structure
to obtain a slab or cluster model.
Without any treatment of the dangling bonds,
the computational model results in an open-shell electronic structure.
Therefore, such models do not reflect the realistic electronic structure.
We previously proposed a Step-by-Step Hydrogen Termination (SSHT) approach 
to circumvent this problem
\cite{Kojima2019_239}.
In this approach,
H atoms are bonded step-by-step to O atoms with
large orbital coefficient values or VCD distributions
until the model exhibits an appropriate energy gap corresponding with experiments.
Notably, in the models thus obtained, 
all the O atoms on the surface are not terminated with H atoms.
Excess hydrogen termination may subsequently lead to excessive electron-doping,
which would yield an unacceptably small energy gap.

In this study, 
we constructed SSHT slab and cluster models 
of the $\gamma$-alumina surface that reproduce the realistic electronic structures.
The infrared (IR) and ultraviolet-visible (UV/Vis) spectra 
were calculated to confirm the reliability of the models.
Thereafter, based on these computational models,
we investigated the regioselectivity of the Cu adsorption on the $\gamma$-alumina surface 
and that of the NO adsorption on the Cu/$\gamma$-alumina surface.
The role of the catalyst support on the NO reduction 
was also discussed using the fragment orbital analysis
\cite{Hoffman1988}
in which molecular orbitals of a system are decomposed into those of its fragments.

\section{\label{SEC2} Theory}
\subsection{Vibronic Coupling Density (VCD)}
Theory of VCD are reviewed in Refs.
\citenum{Sato2009_99} and \citenum{Sato2013_012010}.
Vibronic couplings are the interactions between nuclear vibrations and electrons.
Within a crude adiabatic approximation, 
diagonal vibronic coupling constant (VCC) is defined by
\cite{Fischer1984,Azumi1977_315}
\begin{equation}
	V_{+, \alpha} = 
	\braket{\Psi_+ ({\bf r}; {\bf R}_0) | \mathcal{V}_{\alpha} | \Psi_+ ({\bf r}; {\bf R}_0)},
\end{equation}
where $\ket{\Psi_+ ({\bf r}; {\bf R}_0)}$ is a charge-transfer state that 
depends on electronic coordinates, ${\bf r}$, 
at the equilibrium reference nuclear configuration ${\bf R}_0$.
$\mathcal{V}_{\alpha}$ is the electronic part of a linear vibronic coupling operator
given by
\begin{equation}
	\mathcal{V}_{\alpha} = \left( 
	\frac{\partial \hat{H} ({\bf r}, {\bf R})}{\partial Q_{\alpha}} 
	\right)_{{\bf R}_0},
\end{equation}
where $\hat{H}({\bf r}, {\bf R})$ denotes the molecular Hamiltonian,
${\bf R}$ the nuclear coordinates,
and $Q_{\alpha}$ a mass-weighted normal coordinate of vibrational mode $\alpha$.
Using the Hellmann--Feynman theorem
\cite{Hellmann1937,Feynman1939_340},
the diagonal VCC is expressed as follows:
\begin{equation}
	V_{+, \alpha} = 
	\left( \frac{ \partial E_+ ({\bf R}_0)}{ \partial Q_{\alpha}} \right)_{{\bf R}_0},
\end{equation}
where $E_+ ({\bf R}_0)$ are the eigenvalues of $\ket{\Psi_+ ({\bf r}; {\bf R}_0)}$.
The diagonal VCC can be evaluated from the gradient of a potential energy surface 
with respect to $Q_{\alpha}$ at ${\bf R}_0$.

Owing to the expression of $\mathcal{V}_{\alpha}$ as a sum of one-electron operators, 
$v_{\alpha} (\textit{\textbf{x}})$,
the diagonal VCC is decomposed into the orbital contributions
\cite{Sato2009_99,Sato2013_012010};
that is,
\begin{equation}
	V_{+, \alpha} = \sum_{i \in {\rm occ}} f_{i, \alpha},
\end{equation}
where $f_{i, \alpha}$ is the orbital vibronic coupling constant (OVCC) 
defined as follows:
\cite{Sato2009_99,Sato2013_012010}
\begin{equation}
	f_{i, \alpha} 
	= \braket{\psi_i (\textit{\textbf{x}} )|v_{\alpha} (\textit{\textbf{x}})|\psi_i (\textit{\textbf{x}})}.
\end{equation}
Here, $\ket{\psi_i (\textit{\textbf{x})}}$ represents a one-electron state, 
or molecular orbital,
whereas $\textit{\textbf{x}} = (x,y,z)$ the Cartesian coordinate of a single electron.

The diagonal VCD is given by the spatial distributions of the diagonal VCC
as follows:
\begin{equation}
	V_{+, \alpha} = \int \eta_{+, \alpha} (\textit{\textbf {x}}) \ d \textit{\textbf {x}}.
\end{equation}
$\eta_{+,\alpha}(\textit{\textbf{x}})$ as a function of $\textit{\textbf{x}}$ 
identifies a site where the VCC assumes a large value.
$\eta_{+,\alpha}(\textit{\textbf{x}})$ is divided into the electronic and vibrational terms
as follows:
\begin{equation}
	\eta_{+,\alpha}(\textit{\textbf{x}}) 
	= \Delta \rho(\textit{\textbf{x}}) \times v_{\alpha} (\textit{\textbf{x}}).
\end{equation}
Here, $\Delta \rho (\textit{\textbf{x}})$ is the electron density difference 
between $\ket{\Psi_+ ({\bf r}; {\bf R}_0)}$ and equilibrium reference state,
\begin{equation}
	\Delta \rho(\textit{\textbf{x}}) = \rho_+ (\textit{\textbf{x}}) - \rho_0 (\textit{\textbf{x}}).
\end{equation}
$v_{\alpha} (\textit{\textbf{x}})$ is the potential derivative given by
\begin{equation}
	v_{\alpha} (\textit{\textbf{x}}) 
	= \left( \frac{\partial u (\textit{\textbf{x}})}{\partial Q_{\alpha}} \right)_{{\bf R}_0},
\end{equation}
where $u(\textit{\textbf{x}})$ is an attractive electron-nucleus potential acting on a single electron.
The VCD analysis explores the origin of the VCC from the electronic and vibrational states.

Parr and Yang formulated the frontier orbital theory of chemical reactivity 
in terms of the conceptual density functional theory (DFT)
\cite{Parr1984_4049,Parr1994}.
In their theory, 
the Fukui function is approximately equal to the frontier orbital density, and
the highest occupied molecular orbital (HOMO) density 
is regarded as a reactivity index for electrophilic reactions,
whereas the lowest unoccupied molecular orbital (LUMO) density is for nucleophilic reactions.
Sato \textit{et al.} previously demonstrated that 
the total differential of chemical potential, $d\mu$,
can be expressed using the VCD, $\eta_{\xi}$, along a reaction mode, $\xi$;
\cite{Sato2008_758}
\begin{equation}
	d \mu = 2 \eta_N dN + \int \eta_{\xi} (\textit{\textbf{x}}) \ d\xi d \textit{\textbf{x}},
\end{equation}
where $\eta_N$ denotes the absolute hardness,
$N$ the number of electrons, and
$\eta_{\xi} (\textit{\textbf{x}})$ the VCD of reaction mode $\xi$.
It is possible to conclude that the chemical reactions occur at a site 
where $\eta_{\xi} (\textit{\textbf{x}})$ is the largest
because the chemical reactions are assumed to occur so as to maximize $d\mu$
\cite{Parr1984_4049,Parr1994}.
Therefore, $\eta_{\xi} (\textit{\textbf{x}})$ can be regarded
as a reactivity index for chemical reactions
including the effects of both the electronic and vibrational states on the regioselectivity.
In the current study, 
the reaction mode was selected as an effective mode given by
\cite{Sato2012_257}
\begin{equation}
	d\xi = 
	\sum_{\alpha} \frac{V_{+,\alpha}}{\sqrt{\sum_{\alpha} |V_{+,\alpha}|^2}} dQ_{\alpha},
\end{equation}
which is the steepest descent direction of the structural relaxation after the charge transfer.
$\eta_{\xi} (\textit{\textbf{x}})$ is represented as the product of 
$\Delta \rho (\textit{\textbf{r}})$ and 
potential derivative for the effective mode $v_{\xi} (\textit{\textbf{x}})$.
The VCC with respect to the effective mode quantifies the stabilization 
via the structural relaxation following the charge transfer.

\subsection{Fragment Orbital Analysis}

Divide a molecule into fragments A and B.
A molecular orbital, $\psi_m (\textit{\textbf{x}})$,
is expanded in terms of those of fragments A, $\psi_{k}^{{\rm A}} (\textit{\textbf{x}})$,
and B, $\psi_{l}^{{\rm B}} (\textit{\textbf{x}})$, 
which are known as {\it fragment orbitals}
\cite{Hoffmann1973_7644,Hoffman1988};
\begin{equation}
	\psi_m (\textit{\textbf{x}})
	= \sum_{k} c_{km}^{{\rm A}} \psi_{k}^{{\rm A}} (\textit{\textbf{x}})
	+ \sum_{l} c_{lm}^{{\rm B}} \psi_{l}^{{\rm B}} (\textit{\textbf{x}}),
\end{equation}
where $k$ and $l$ run over the molecular orbitals of A and B, respectively.
The contribution of $\psi_{k}^{{\rm A}} (\textit{\textbf{x}})$ to 
$\psi_m (\textit{\textbf{x}})$, $P_{km}$, is calculated from
\begin{equation}
	P_{km} 
	= |c_{km}|^2 
	+ \sum_l c_{km}^{{\rm A}} c_{lm}^{{\rm B}} 
	\braket{\psi_k^{{\rm A}}(\textit{\textbf{x}})|\psi_l^{{\rm B}}(\textit{\textbf{x}})}.
	\label{Eq:frag}
\end{equation}
$P_{km}$ represents the proportion of 
$\psi_{k}^{{\rm A}} (\textit{\textbf{x}})$ in $\psi_m (\textit{\textbf{x}})$
because $\sum_{k} P_{km} + \sum_{l} P_{lm} = 1$.

\section{\label{SEC3}Methods of Calculations}

\begin{figure}[!ht]
\centering
\includegraphics[width=0.9\hsize]{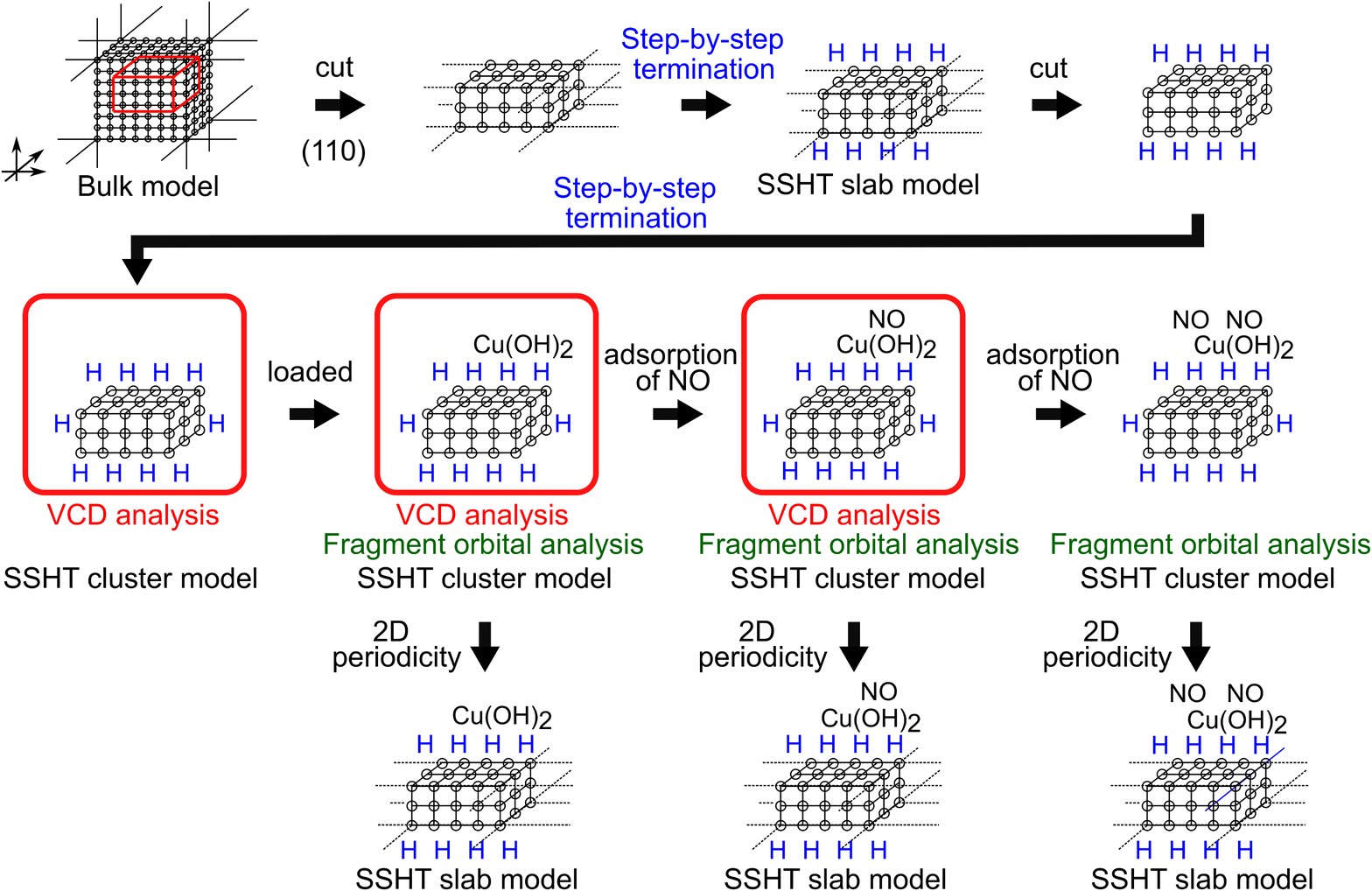}
\caption{
Schematic diagram of the computational procedure.
The 3D bulk model is obtained from Refs.
\citenum{Krokidis2001_5121,Digne2002_1,Digne2004_54}.
}
\label{FIG1}
\end{figure}

Figure~\ref{FIG1} illustrates the computational procedure.
$\gamma$-alumina is a metastable phase
generated during a thermal transition from a Bohemito to $\alpha$-alumina.
Raybaud \textit{et al.} constructed a 3D bulk model of the $\gamma$-alumina
by following the transition process using molecular dynamics simulations
\cite{Krokidis2001_5121,Digne2002_1,Digne2004_54}.
Furthermore, they estimated that the (110), (100), and (111) surfaces respectively occupy 
74, 16, and 10$\%$ of the total area of the $\gamma$-alumina surface
\cite{Digne2002_1,Digne2004_54},
which is consistent with experiments
\cite{Nortier1990_141}.
Therefore, we focused on the (110) surface
because its surface area 
is dominant.

The slab model of the (110) surface was obtained by cutting the 3D bulk model
\cite{Krokidis2001_5121,Digne2002_1,Digne2004_54}.
The dangling bonds resulting from the homolytic cleavage of O atoms 
were treated using the SSHT approach.
These dangling bonds should be terminated with H atoms 
because hydrations occur in the production of the $\gamma$-alumina
\cite{Morterra1996_497}.
In fact, the experimental IR spectra indicate
the existence of OH groups on the $\gamma$-alumina surface
\cite{Busca1993_1492,Tsyganenko1996_4843,Pecharroman1999_6160,Paglia2004_1914}.
A cluster model was also obtained by cutting the slab model.
Compared with a slab model, 
a cluster model has the following advantages:
(a) a single model contains multiple surfaces,
(b) the computational cost is low
because of the lack of ${\bf k}$ points in band calculations, 
and
(c) the calculations of ionic states are possible.
The dangling bonds generated when cutting the slab model 
were again treated using the SSHT approach.

The VCD analysis for the cluster model was performed
to identify the Cu$^{2+}$ adsorption site.
Cu(OH)$_2$ is loaded on the $\gamma$-alumina surface
because experiments indicate that 
a highly dispersed Cu$^{2+}$ species has a coordination environment similar to Cu(OH)$_2$.
\cite{Yamamoto2002_2449}.
The adsorption sites of NO on the Cu/$\gamma$-alumina surface were also
determined based on the VCD analyses for the cluster models.
The adsorbed structures were obtained via geometry optimizations 
using the cluster and slab models.
The optimizations were started from the structure 
in which adsorbents were placed on the sites where the VCD was localized.
The fragment orbital analyses for the cluster models were performed
to clarify the role of the $\gamma$-alumina support on the NO reduction.

The geometries of the neutral cluster models were optimized.
The forces acting on the nuclei in the monocationic states 
at the neutral optimized structures were calculated to evaluate the VCC and VCD.
The monocationic states were regarded as the charge-transfer states
because the electron donations from the $\gamma$-alumina to Cu as well as 
that from the Cu/$\gamma$-alumina to NO were expected.
The calculations of the cluster models were performed using Gaussian 09 
\cite{Frisch2013D,Frisch2013E}
at the B3LYP/3-21G level of theory. 
In the OVCC calculations,
the restricted open-shell Kohn--Sham calculation was employed for simplicity.
The IR spectrum was computed
by summing the IR intensities broadened with the Gaussian function.
The UV/Vis spectrum was computed 
by summing the oscillator strengths,
which were evaluated by the time-dependent DFT theory, 
broadened with the Gaussian function.
The fragment orbital analyses were performed based on the extended H\"uckel theory 
\cite{Hoffmann1963_1397}
with parameters provided in YAeHMOP
\cite{yaehmop}.
The VCC, VCD, and fragment orbital analyses were performed 
using our in-house codes.
The geometries of the slab models were also optimized.
The calculations of the slab models were performed
using Amsterdam Density Functional Band Structure Package 
\cite{Velde1991_7888,Velde1992_84}
with the local density approximation and double zeta basis set. 

\section{Results and Discussion}

\subsection{Model Building of the $\gamma$-Alumina Surface.}
\begin{figure}[!ht]
\centering
\includegraphics[width=0.8\hsize]{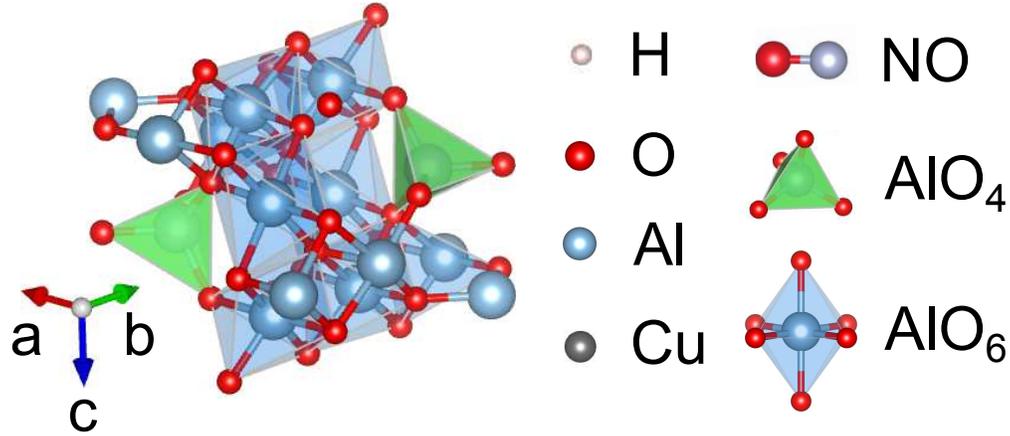}
\caption{
Unit cell of the $\gamma$-alumina (110) surface slab 
obtained by cutting the 3D bulk model in Refs.
\citenum{Krokidis2001_5121} and \citenum{Digne2004_54}.
The lattice constants are $a=8.068$ and $b=8.413$ \AA.
The unit cell has an inversion center.
Hereafter, the atomic symbols in this figure are used without mentioning them.
}
\label{FIG2}
\end{figure}

Figure~\ref{FIG2} illustrates a unit cell of the $\gamma$-alumina (110) surface slab obtained
by cutting the three-dimensional bulk model
\cite{Krokidis2001_5121,Digne2002_1,Digne2004_54}.
The slab has a thickness of one unit cell layer in the $z$-direction.
The lattice constants in the $x$- and $y$-directions are $a=8.068$ and $b=8.413$ \AA, respectively.
The unit cell has an inversion center.
The $\gamma$-alumina consists of Al atoms coordinated by four or six O atoms,
and the chemical formula of the unit cell is Al$_{16}$O$_{32}$.

\begin{figure}[!ht]
\centering
\includegraphics[width=0.6\hsize]{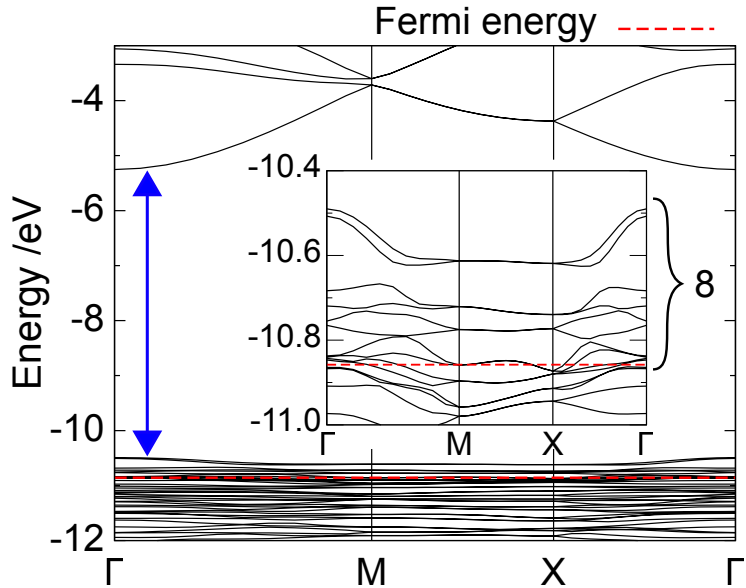}
\caption{
Calculated band structure of the $\gamma$-alumina slab without hydrogen termination.
Sixteen H atoms should be bonded to the O atoms 
for the slab model to have an appropriate band gap indicated by a blue vertical arrow.
}
\label{FIG3}
\end{figure}

Figure~\ref{FIG3} presents the calculated band structure of the slab 
without hydrogen termination.
The band gap of the $\gamma$-alumina surface was
experimentally observed ranging from 2.5 to 8.7 eV
\cite{Ealet1994_92}.
However, the calculated band gap of the bare slab is zero
because the Fermi level intersects with the occupied bands.
This discrepancy is attributed to 
the dangling bonds generated due to the homolytic cleavage of the O atoms
when cutting the 3D crystal structure.
The eight virtual bands should be occupied 
for the slab model to have an appropriate band gap corresponding with experiments.
The hydrogen termination involves the electron-doping.
Suppose that the formal charges of Al, O, and H are respectively $3+$, $2-$, and $1+$,
the net charge of the slab model becomes zero after introducing 16 H atoms
because the chemical formula of the obtained unit cell becomes Al$_{16}$O$_{32}$H$_{16}$.
In other words, the hydrogen termination results in 
the appropriate band gap and oxidation state.

\begin{figure}[!ht]
\centering
\includegraphics[width=0.9\hsize]{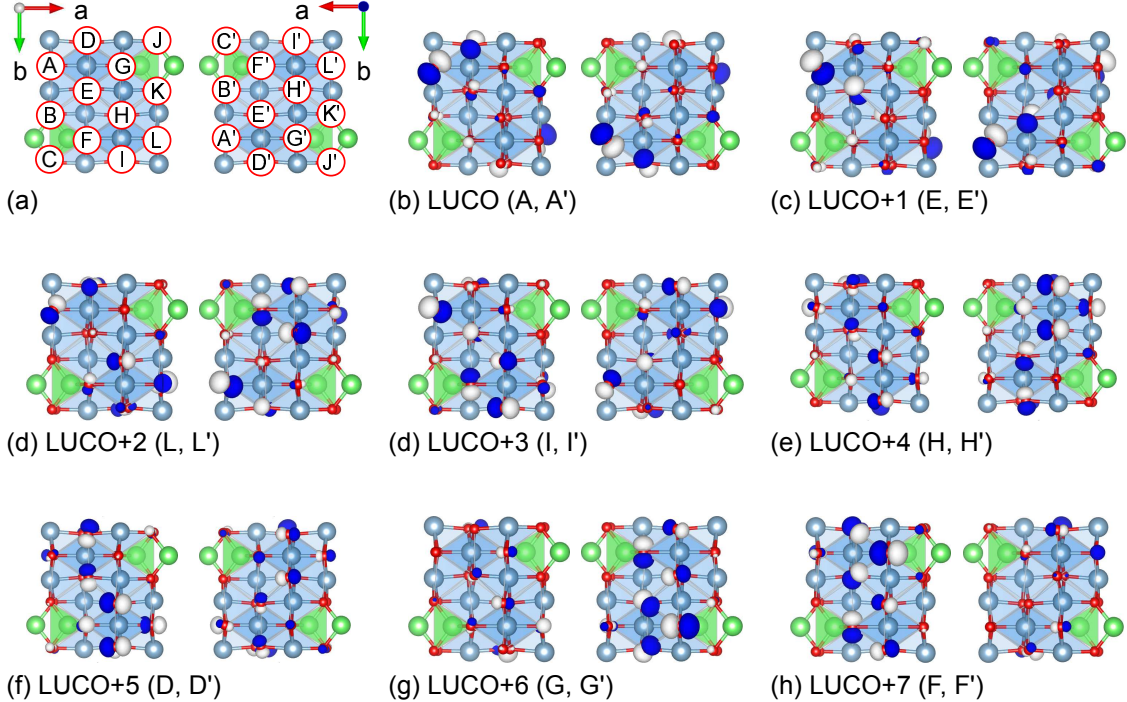}
\caption{
(a)
Labels of O atoms on the (110) surface,
where the equivalent O atoms with respect to an inversion center 
are labeled using similar characters such as A and A'.
(b)--(h)
LUCO--LUCO+7 at the $\Gamma$-point of the bare slab.
The O atoms that should be terminated with H atoms 
because of the large crystal orbital coefficients
are provided in parentheses.
Isosurface values are $5.0\times10^{-2}$ a.u.
}
\label{FIG4}
\end{figure}

\begin{figure}[!ht]
\centering
\includegraphics[width=0.9\hsize]{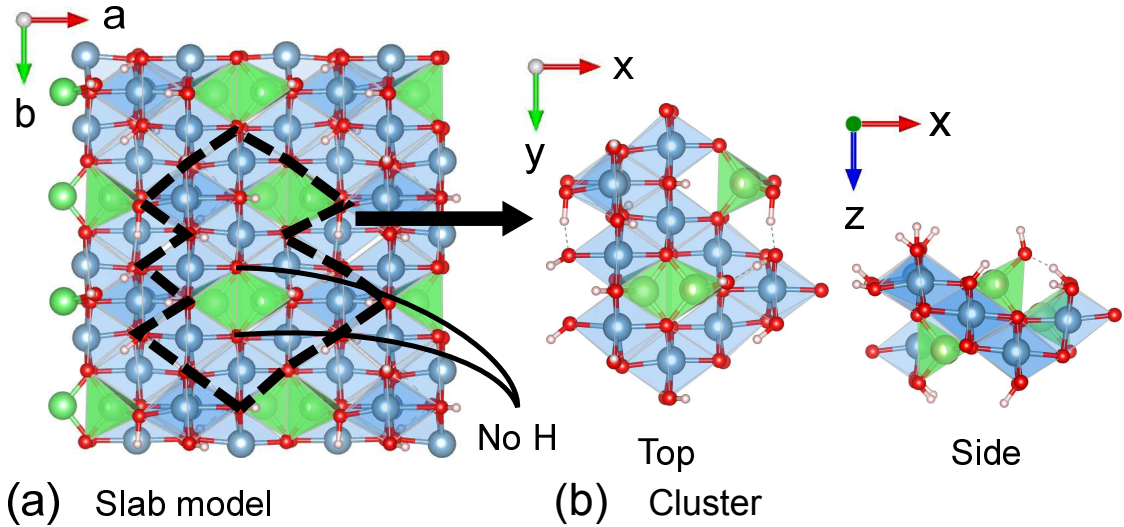}
\caption{
(a) SSHT slab model of the $\gamma$-alumina (110) surface
after the geometry optimization.
(b) Bare cluster obtained by
cutting the slab model along with the dashed lines.
The symmetry of the bare cluster is $C_1$.
}
\label{FIG5}
\end{figure}

The sites for the hydrogen termination are determined 
based on the crystal orbital coefficients of the virtual bands at the $\Gamma$-point.
Figure~\ref{FIG4} (a) shows the O atoms on the front and reverse sides of the (110) surface
where the equivalent O atoms with respect to an inversion center are labeled using similar characters
such as A and A'.
For each virtual band,
two H atoms are bonded stepwise to the equivalent O atoms with large coefficients.
Figure~\ref{FIG4} (b)--(h) shows the 
unoccupied crystal orbitals at the $\Gamma$-point of the bare slab.
The lowest unoccupied crystal orbital (LUCO) is mainly distributed on the O atoms at A and A'.
Hence, two H atoms are bonded to these O atoms.
The LUCO+1 is distributed at A, A', E, and E'.
Since the O atoms at A and A' are already terminated,
two H atoms are bonded to the atoms at E and E'.
In a similar manner, 
H atoms are bonded to the O atoms at L, I, H, D, G, and F 
based on the coefficients of 
the LUCO+2, +3, +4, +5, +6, and +7, respectively.
Notably, the O atoms at the vertex of the tetrahedral Al species, 
i.e., B, C, J, and K, are not terminated
because the coefficients of these O atoms are small.
Figure~\ref{FIG5} (a) illustrates the SSHT slab model 
after the geometry optimization of the model obtained above.
This model has an appropriate wide band gap of 3.97 eV corresponding with experiments
(Fig. S1).
Cartesian coordinates of the SSHT slab model are tabulated in Table~S1.

\begin{figure}[!ht]
\centering
\includegraphics[width=0.5\hsize]{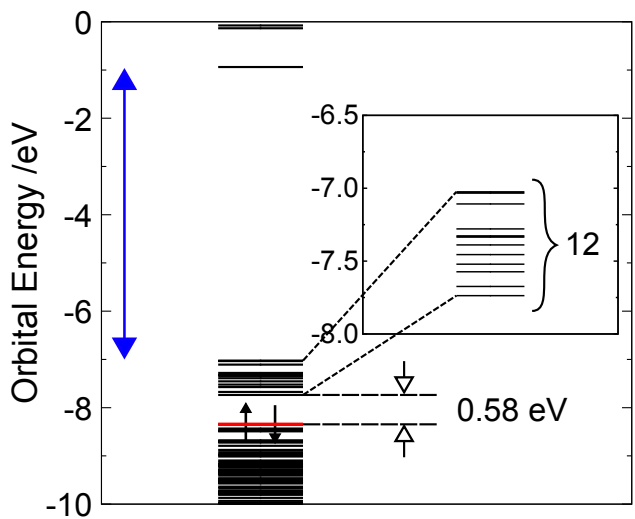}
\caption{
Calculated orbital levels of the bare cluster.
Twenty-four H atoms should be bonded to O atoms 
for the cluster model to have an appropriate energy gap 
indicated by a blue vertical arrow.
}
\label{FIG6}
\end{figure}

A cluster model of the $\gamma$-alumina surface is 
obtained by cutting the SSHT slab model 
so that the O atoms without the hydrogen termination are centered
(Fig.~\ref{FIG5} (b)).
The chemical formula of the bare cluster is Al$_{13}$O$_{40}$H$_{17}$.
The symmetry is $C_1$.
Figure~\ref{FIG6} presents the orbital levels of the bare cluster.
The energy gap of the bare cluster is calculated to be 0.58 eV,
which is smaller than the experimental values of 2.5$\sim$8.7 eV
\cite{Ealet1994_92}.
This is because dangling bonds are generated when cutting the slab model.
The twelve virtual orbitals are required to be fully occupied 
for the cluster model to have an appropriate energy gap corresponding with experiments.
The addition of twenty-four H atoms results in the 
appropriate energy gap as well as the oxidation state
because the chemical formula of the obtained cluster model 
becomes Al$_{13}$O$_{40}$H$_{41}$.
Figure S2 (a) shows the labels of the O atoms of the bare cluster
in which no equivalent O atom exists due to asymmetry.
For the bare cluster, 
H atom is already bonded to 
the O atoms at A, B, C, D, F, G, H, J, M, P, R, S, T, d, f, g, and p.
The sites for further hydrogen termination 
are determined based on the molecular orbital coefficients.
Two H atoms are bonded to the O atoms at E and o 
because of the large coefficients of the LUMO
(Fig.~S2 (b)).
In a similar manner,
two H atoms are bonded to the O atoms at 
C and g, i and r, M and h, j and o, Q and m, I and q, 
I and N, R and r, T and U, A and a, and s and t
based on the coefficients of the 
LUMO+1, +2, +3, +4, +5, +6, +7, +8, +9, +10, and +11, respectively
(Fig.~S2 (c)--(m)).
The O atoms at the vertex of the tetrahedral Al species, 
i.e. K, L, O, k, $\ell$,
are not terminated 
because the coefficients of theses O atoms are small,
which is the same situation as in the slab model.

\begin{figure}[!ht]
\centering
\includegraphics[width=0.6\hsize]{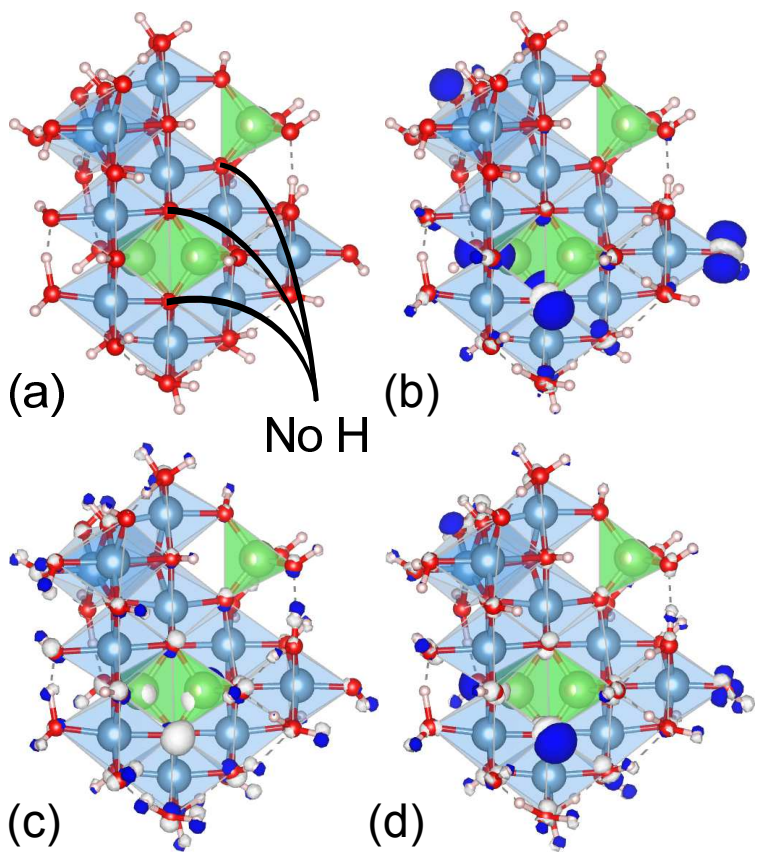}
\caption{
(a) SSHT cluster model of the $\gamma$-alumina (110) surface
after the geometry optimization,
and its
(b) electron density difference, $\Delta \rho(\textit{\textbf{x}})$, 
(c) potential derivative, $v_{\xi} (\textit{\textbf{x}})$, and
(d) VCD, $\eta_{\xi} (\textit{\textbf{x}})$.
Isosurface values of $\Delta \rho(\textit{\textbf{x}})$, 
$v_{\xi} (\textit{\textbf{x}})$, and 
$\eta_{\xi} (\textit{\textbf{x}})$
are $3.0\times10^{-3}$, 
$7.0\times10^{-3}$, and 
$2.0\times10^{-5}$ a.u., respectively.
}
\label{FIG7}
\end{figure}

Figure~\ref{FIG7} (a) illustrates the SSHT cluster model 
after the geometry optimization of the cluster model obtained above.
This model has an appropriate wide energy gap of 5.4 eV corresponding with experiments (Fig.~S3). 
The O atoms at A, C, I, M, S, T, g, o, and r are terminated with two H atoms, 
i.e., H$_2$O is adsorbed on the boundary surface.
Therefore, in contrast to the slab model,
hydrogen bonds exist at the edge of the cluster model.
Cartesian coordinates of the SSHT cluster model are tabulated in Table~S2.

\subsection{Regioselectivity of the Cu and NO Adsorption.}

The VCD analysis for the $\gamma$-alumina cluster model was performed 
to identify a Cu$^{2+}$ adsorption site
(Figs.~\ref{FIG6}~(b)--(d)).
$\Delta \rho(\textit{\textbf{x}})$ is distributed on a few O atoms 
located at the center and edge of the model.
Therefore, the adsorption sites are ambiguous based on the frontier orbital theory.
In contrast,
$v_{\xi}(\textit{\textbf{x}})$ also has large values on the central O atom.
Consequently, $\eta_{\xi}(\textit{\textbf{x}})$,
which is given by the product of  
$\Delta \rho(\textit{\textbf{x}})$ and $v_{\xi}(\textit{\textbf{x}})$,
is localized on the central O atom at the vertex of the tetrahedral Al species.
Therefore, 
$\eta_{\xi}(\textit{\textbf{x}})$ clarifies the regioselectivity 
more clearly than $\Delta \rho(\textit{\textbf{x}})$.
Notably, the O atom where $\eta_{\xi}(\textit{\textbf{x}})$ is localized 
is not terminated with H atoms.
The O atom without a hydrogen termination has a lone electron pair
and acts as a Lewis base
\cite{Trueba2005_3393}.
Thus, this site can be the adsorption site of Cu(OH)$_2$.

The OVCCs were calculated to determine
molecular orbitals mainly contributing to the interactions 
of the surface with Cu(OH)$_2$ (Table~S1).
The OVCC of the HOMO has the largest value,
which suggests that the HOMO mainly contributes to
the reaction with Cu(OH)$_2$.
This is because the HOMO and its orbital vibronic coupling density (OVCD), 
the density form of the OVCC, 
have large values on the Lewis-basic O atom (Figs.~S4 (a) and (b)).
The second and third largest OVCCs are for the HOMO-2 and HOMO-1, respectively.
These orbitals are also distributed on the Lewis-basic site 
(Figs.~S4 (c)--(f)),
and moderately contribute to the reaction with Cu(OH)$_2$.

\begin{figure}[!ht]
\centering
\includegraphics[width=0.8\hsize]{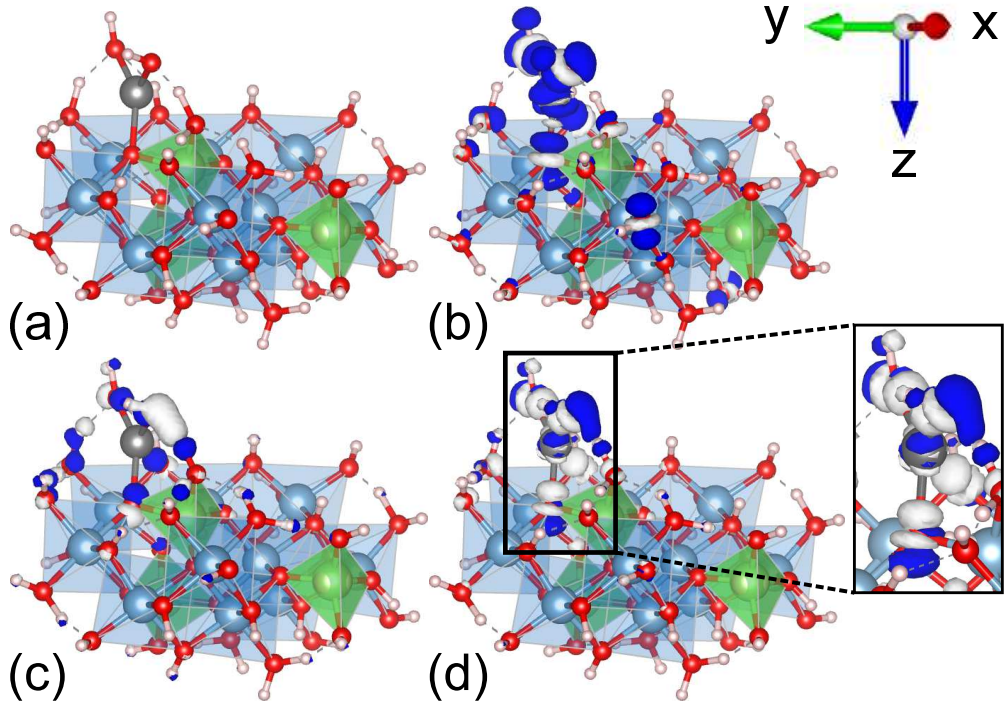}
\caption{
(a) Geometry-optimized structure of the Cu(OH)$_2$/$\gamma$-alumina cluster model,
and its
(b) electron density difference, $\Delta \rho(\textit{\textbf{x}})$,
(c) potential derivative, $v_{\xi}(\textit{\textbf{x}})$, and
(d) VCD, $\eta_{\xi}(\textit{\textbf{x}})$.
Isosurface values of $\Delta \rho(\textit{\textbf{x}})$, 
$v_{\xi} (\textit{\textbf{x}})$, and 
$\eta_{\xi} (\textit{\textbf{x}})$
are $3.0\times10^{-3}$, 
$7.0\times10^{-3}$, and 
$2.0\times10^{-5}$ a.u., respectively.
}
\label{FIG8}
\end{figure}

Figure~\ref{FIG8} (a) illustrates the geometry-optimized structure 
of the Cu(OH)$_2$/$\gamma$-alumina model predicted by the above VCD analysis.
The optimization exhibits that Cu(OH)$_2$ adsorbs on the Lewis-basic O atom.
The VCD analysis for the Cu(OH)$_2$/$\gamma$-alumina model was performed 
to identify an NO adsorption site
(Figs~\ref{FIG8} (b)--(d)).
Since 
NO is predicted to be adsorb on 
the sites where the VCD is localized in Cu(OH)$_2$
because 
$\Delta \rho(\textit{\textbf{x}})$, 
$v_{\xi}(\textit{\textbf{x}})$, and
$\eta_{\xi}(\textit{\textbf{x}})$ are localized on Cu(OH)$_2$.
$\eta_{\xi}(\textit{\textbf{x}})$ is distributed
not only on Cu(OH)$_2$ 
but also on the Lewis-basic O atom of the $\gamma$-alumina surface.
This indicates that
the excessive reaction energies arising from the NO adsorption
dissipate to the $\gamma$-alumina surface.
In other words, the catalyst support can be regarded as a heat bath.
In addition,
Cu(OH)$_2$ can be tightly bonded on the $\gamma$-alumina surface
because of the localization of $\eta_{\xi}(\textit{\textbf{x}})$ 
on the Lewis-basic site,
which results in the highly dispersed Cu$^{2+}$ species.
This is so-called anchoring effect.

\begin{figure}[!ht]
\centering
\includegraphics[width=0.7\hsize]{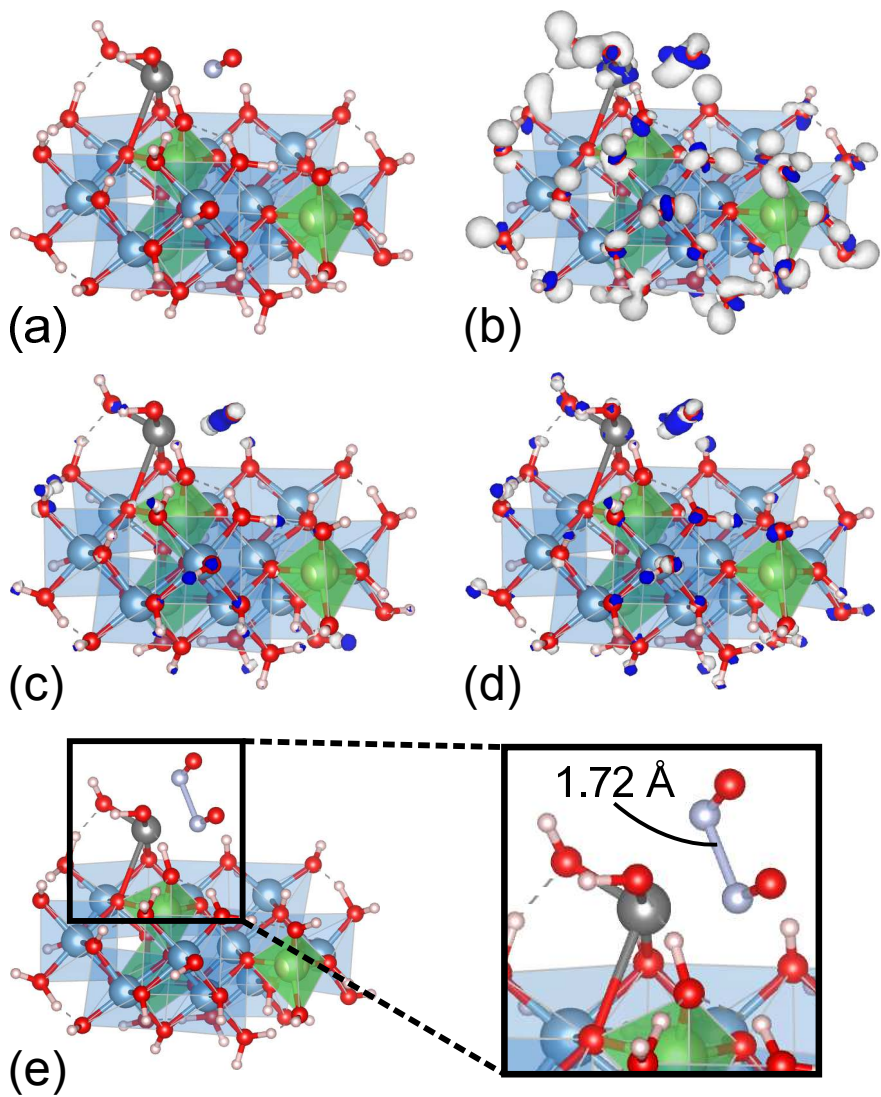}
\caption{
 (a) Geometry-optimized structure of the NO-adsorbed 
Cu(OH)$_2$/$\gamma$-alumina cluster model,
and its
(b) electron density difference, $\Delta \rho(\textit{\textbf{x}})$,
(c) potential derivative, $v_{\xi}(\textit{\textbf{x}})$, and
(d) VCD, $\eta_{\xi}(\textit{\textbf{x}})$.
Isosurface values of $\Delta \rho(\textit{\textbf{x}})$, 
$v_{\xi} (\textit{\textbf{x}})$, and 
$\eta_{\xi} (\textit{\textbf{x}})$
are $7.0\times10^{-2}$, $1.0\times10^{-2}$, and $5.0\times10^{-4}$ a.u., respectively.
(e) Geometry-optimized structure of the Cu(OH)$_2$/$\gamma$-alumina model 
with the two NO molecules adsorbed.
}
  \label{FIG9}
\end{figure}

Figure~\ref{FIG9} (a) illustrates the geometry optimized structure of 
the NO-adsorbed Cu(OH)$_2$/$\gamma$-alumina model
where the geometry optimization was started 
from the structure in which NO is placed
on the site where the VCD is localized.
To manifest the validity of the VCD as a reactivity index,
the geometry optimizations were also performed 
where the initial NO positions are changed
on the Cu(OH)$_2$/$\gamma$-alumina model.
Consequently, 
the optimized structure predicted by the VCD analysis
is discovered to be the most energetically stable (Fig.~S5).
Therefore, the regioselectivity of the NO adsorption in the present model 
is successfully reproduced via the VCD analysis.

The VCD analysis of the NO-adsorbed Cu(OH)$_2$/$\gamma$-alumina model was performed 
to identify a second NO adsorption site
(Figs~\ref{FIG9} (b)--(d)).
$\Delta \rho(\textit{\textbf{x}})$ is delocalized 
over the NO, Cu(OH)$_2$, and $\gamma$-alumina.
Thus, the site for the second NO adsorption is ambiguous
as long as only the electronic state is considered.
In contrast, $v_{\xi}(\textit{\textbf{x}})$ is mainly distributed on the first NO.
Therefore, $\eta_{\xi}(\textit{\textbf{x}})$ is localized on the NO 
that is previously adsorbed on the Cu(OH)$_2$/$\gamma$-alumina model,
and the second NO is predicted to be adsorbed towards the first NO.
Again, $\eta_{\xi}(\textit{\textbf{x}})$ clarifies the regioselectivity 
more clearly than $\Delta \rho(\textit{\textbf{x}})$.
Figure~\ref{FIG9} (d) illustrates the geometry-optimized structure of  
the Cu(OH)$_2$/$\gamma$-alumina model with the two NO molecules adsorbed.
The NO molecules undergo dimerization on Cu(OH)$_2$.
The distance between the N atoms is 1.72 $\AA$,
which is sufficiently close to form a N--N bond.

\begin{figure}[!ht]
\centering
\includegraphics[width=0.55\hsize]{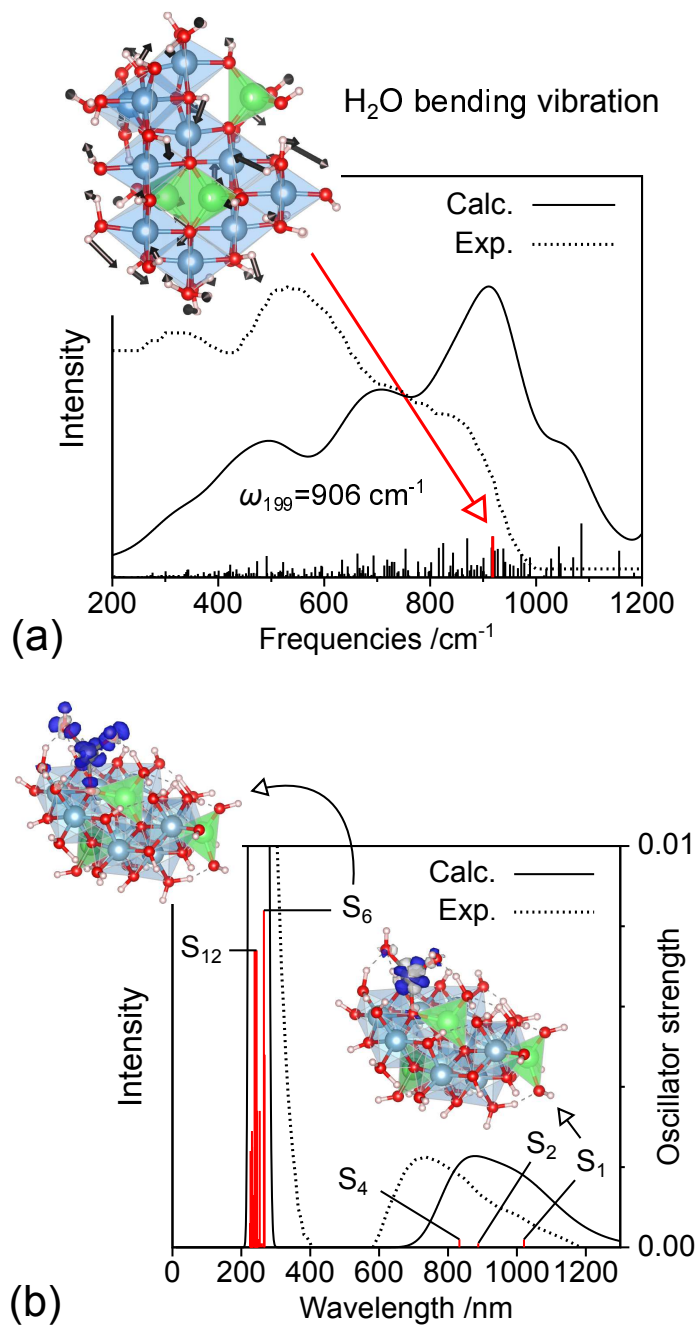}
\caption{
(a) 
Simulated IR spectrum of the $\gamma$-alumina cluster model
where the linewidth of the Gaussian function is 40 cm$^{-1}$.
Inset shows vibrational mode 199.
Experimental IR spectrum treated at 600$^{\circ}$C 
is obtained from Ref.~\citenum{Pecharroman1999_6160}.
(b) 
Simulated UV/Vis spectrum of the Cu(OH)$_2$/$\gamma$-alumina cluster model
with the excited electronic states from S$_1$ to S$_{25}$ considered.
The linewidth of the Gaussian function is 1000 cm$^{-1}$.
Inset shows the electron density difference 
between S$_1$ and S$_0$ as well as S$_6$ and S$_0$.
Experimental UV/Vis spectrum is obtained from Ref.~\citenum{Yamamoto2002_113}.
}
\label{FIG10}
\end{figure}

The IR spectrum of the $\gamma$-alumina cluster model as well as 
the UV/Vis spectrum of the Cu(OH)$_2$/$\gamma$-alumina cluster model
were calculated to check the reliability of the models discussed thus far.
Figure~\ref{FIG10} (a) shows the simulated and experimental IR spectra.
The $\gamma$-alumina cluster model is considered to be realistic
because the lineshape of the simulated IR spectrum 
appropriately reproduces that of the experimental spectrum. 
The observed IR spectrum in the ranging from 200 to 1200 cm$^{-1}$ 
is assigned to the H$_2$O bending vibrations based on the calculation.
Figure~\ref{FIG10} (b) shows the simulated and experimental UV/Vis spectrum.
The lineshape of the simulated UV/Vis spectrum 
also appropriately reproduces that of the experimental spectrum,
and this indicates that the Cu(OH)$_2$/$\gamma$-alumina cluster model is realistic.
The observed spectrum ranging from 600 to 1200 nm is attributed 
to the absorption of S$_1$, S$_2$, and S$_4$.
The electron density difference between S$_1$ and S$_0$ is localized on the Cu atom
with the $d$ orbital distributions.
Similar electron density differences are obtained
between S$_2$ and S$_0$ as well as S$_4$ and S$_0$ (Fig.~S6 (a)--(c)).
Therefore, the absorption in this region 
is assigned to the $d$-$d$ transitions of the Cu atom.
The observed spectrum below 400 nm 
is mainly attributed to the absorption of S$_6$ and S$_{12}$.
The electron density differences between S$_6$ and S$_0$ as well as S$_{12}$ and S$_{0}$ 
are distributed over Cu(OH)$_2$ (Figs.~S6 (d) and (e)).

The geometry-optimized structures were also obtained using the slab models
(Figs.~S7 (a)--(c)).
These optimized structures are in good agreement with 
those using the cluster models.
Therefore, the cluster models appropriately reproduce
the Cu and NO adsorptions on the slab models.

It has been reported by experiments that
the tetrahedral Al species transform the structure by the Cu loading
\cite{Hosokawa2017_74}.
Hence, the interatomic distance and Mulliken charge of the tetrahedral Al species
on which Cu(OH)$_2$ is adsorbed were computed
(Fig.~S8 and Table~S4).
As a result of the Cu adsorption
on the $\gamma$-alumina cluster model,
the distance between Al and O bonded to Cu is increased 
from 1.757 to 1.823 $\AA$,
whereas the other three Al--O distances remain unchanged.
In addition,
the Mulliken charge of Al is increased from +0.928 to +1.043.
This trend is the same in the slab model
(Table~S4).
These results indicate that
one of the Al--O distances in the tetrahedral Al species is increased 
by the electron transfer from the $\gamma$-alumina to Cu.
To explain this behavior,
the orbital levels of the geometry-optimized AlO$_4$ with the $T_d$ symmetry were calculated 
(Fig.~S9).
Five electrons should be given to AlO$_4$ 
for satisfying the appropriate oxidation number
because the formal charge of Al is $3+$ and of O is $2-$,
The HOMO with the $T_1$ symmetry is the three-fold degenerate.
The symmetrical product of $T_1$ is determined as follows:
\begin{equation}
	[T_1^2] = A_1 \oplus E \oplus T_2.
\end{equation}
In addition,
the VCC of the $t_2$ modes are the largest
except for the totally-symmetric $a_1$ mode
(Table~S5).
Thus, the electron drawing from the HOMO induces 
the $T \otimes t_2$ Jahn--Teller distortion
\cite{Bersuker2006}.
The Jahn--Teller distortion in the $t_2$ modes 
lowers the symmetry of the tetrahedron to $C_{3v}$
\cite{Ceulemans1989_125},
which rationalizes the elongation of one of the Al--O bonds 
in the tetrahedral Al species upon Cu adsorption.

\subsection{
Orbital Levels of the 
NO- and O$_2$-Adsorbed Cu/$\gamma$-Alumina.
}

\begin{figure}[!ht]
\centering
\includegraphics[width=0.90\hsize]{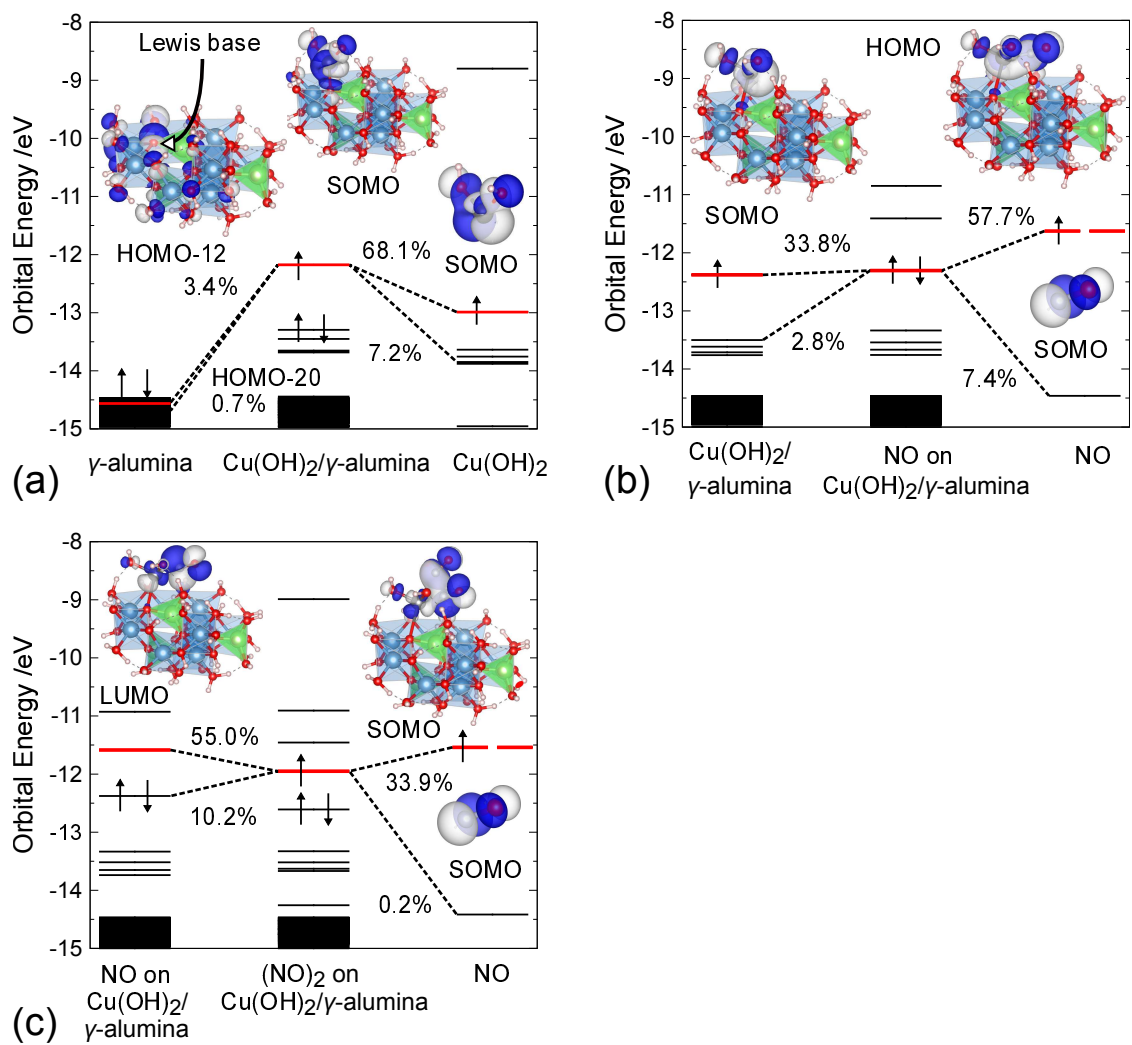}
\caption{
Fragment orbital analyses of the
(a) Cu(OH)$_2$/$\gamma$-alumina,
(b) NO-adsorbed Cu(OH)$_2$/$\gamma$-alumina, and
(c) (NO)$_2$-adsorbed Cu(OH)$_2$/$\gamma$-alumina cluster models.
Inset shows the molecular orbitals.
Values of $P_{km}$ are provided in percentages.
}
\label{FIG11}
\end{figure}

The fragment orbital analyses were performed within the extended H\"uckel theory
to investigate the role of the catalyst and its support.
Figure~\ref{FIG11} (a) illustrates the decomposition of
the orbitals of the Cu(OH)$_2$/$\gamma$-alumina cluster model
into those of Cu(OH)$_2$ and $\gamma$-alumina fragments.
The singly occupied molecular orbital (SOMO) of the Cu(OH)$_2$/$\gamma$-alumina 
is raised with respect to that of the isolated Cu(OH)$_2$
because of the orbital interactions of the isolated Cu(OH)$_2$
with the $\gamma$-alumina.
The main components of the SOMO of the Cu(OH)$_2$/$\gamma$-alumina are
the SOMO of the Cu(OH)$_2$ ($P_{km}$=68.1\%) and the HOMO-12 of the $\gamma$-alumina (3.4\%).
The orbitals exhibiting the second largest contribution in the $\gamma$-alumina 
is the HOMO-20 (0.7\%), 
whose ratio is significantly smaller than the HOMO-12.
Thus,
the HOMO-12 mainly contributes to the interactions with Cu(OH)$_2$.
This is because the HOMO-12 have large values on the Lewis-basic O atom.

Figure~\ref{FIG11} (b) illustrates the decomposition of
the orbitals of the NO-adsorbed Cu(OH)$_2$/$\gamma$-alumina model
into those of NO and Cu(OH)$_2$/$\gamma$-alumina fragments.
The HOMO of the NO-adsorbed Cu(OH)$_2$/$\gamma$-alumina,
which is distributed over the Cu(OH)$_2$ and NO, 
mainly consists of the SOMOs of the Cu(OH)$_2$/$\gamma$-alumina and NO.
Since the SOMO of the Cu(OH)$_2$/$\gamma$-alumina is energetically close to that of NO,
the electron transfer between them can occur.
This is because the $\gamma$-alumina support raises the SOMO of the Cu(OH)$_2$.
Therefore, the role of the catalyst support 
is to shift the frontier orbital levels of the catalyst.
In addition, 
the $\gamma$-alumina hardly interacts with NO
because the occupied orbitals of the $\gamma$-alumina are significantly lower than the SOMO of NO.
This corresponds with the experimental observations that
the $\gamma$-alumina itself exhibits the low catalytic activity for the NO reduction
\cite{Yamamoto2002_2449}.

Figure~\ref{FIG11} (c) illustrates the decomposition of
the orbitals of the (NO)$_2$-adsorbed Cu(OH)$_2$/$\gamma$-alumina model
into those of NO and NO-adsorbed Cu(OH)$_2$/$\gamma$-alumina fragments.
The LUMO of the NO-adsorbed Cu(OH)$_2$/$\gamma$-alumina mainly interacts with the SOMO of NO.
The SOMO of the (NO)$_2$-adsorbed Cu(OH)$_2$/$\gamma$-alumina
has the N--N bonding and N--O anti-bonding,
which is a characteristic of the (NO)$_2$ 2$\pi$ orbital.
The Mulliken charge of (NO)$_2$ on the Cu(OH)$_2$/$\gamma$-alumina
was calculated to be -0.26.
Therefore, the electron back-donation occurs 
from the Cu(OH)$_2$/$\gamma$-alumina to (NO)$_2$.
This results in the occupation of the (NO)$_2$ 2$\pi$ orbital,
which can be the driving force of the NO reduction to N$_2$.

Finally,
the orbital levels of the NO- and O$_2$-adsorbed Cu(OH)$_2$/$\gamma$-alumina were compared
to show the ease of the NO adsorption on the Cu/$\gamma$-alumina under the oxidative condition.
The geometries of the O$_2$-adsorbed Cu(OH)$_2$/$\gamma$-alumina 
were optimized using the cluster and slab models, 
which have similar structures where O$_2$ is adsorbed on Cu(OH)$_2$
(Figs~S10 (a) and (b)).
Figures~S11 (a) and (b) present the results of the fragment orbital analyses 
of the NO- and O$_2$-adsorbed Cu(OH)$_2$/$\gamma$-alumina cluster models.
The HOMO of the NO-adsorbed Cu(OH)$_2$/$\gamma$-alumina has the bonding character
between Cu and NO,
whereas the SOMO of the O$_2$-adsorbed Cu(OH)$_2$/$\gamma$-alumina 
has the anti-bonding character between Cu and O$_2$.
This difference appears in the interatomic distance between Cu and NO or O$_2$,
where the calculated Cu--N bond (1.775 \AA) is shorter than the Cu--O bond (1.963 \AA).
The bond order between atom X and Y,
which quantifies the strength of bond,
is defined as follows:
\begin{equation}
	\sum_{\mu \in {\rm X}} \sum_{\nu \in {\rm Y}} \sum_i
	n_i C_{\mu i}^* C_{\nu i} S_{\mu \nu},
\end{equation}
where
$n_i$ denotes the electron occupation number of molecular orbital $i$,
$C_{\mu i}$ the $i$-th molecular orbital coefficient
of atomic orbital $\mu$ belonging to X, and
S$_{\mu \nu}$ the overlap integral between $\mu$ and $\nu$.
The bond orders of the Cu--N and Cu--O bonds were respectively
computed to be 0.2741 and 0.1144,
which indicates that the Cu--N bond is stronger than the Cu--O bond.
This suggests that
the curvature of the potential energy surface 
of the O$_2$-adsorbed Cu(OH)$_2$/$\gamma$-alumina
with respect to the Cu-O distance 
is larger than that of the NO-adsorbed Cu(OH)$_2$/$\gamma$-alumina
with respect to the Cu-N distance.
In other words, 
the Cu--O bond is considered to be weakly bounded 
due to the small force constant
and, therefore, easy to dissociate.
Notably,
the stabilization energy resulting from the orbital interactions 
of the Cu(OH)$_2$/$\gamma$-alumina with NO (-2.730 eV) is
comparable to that with O$_2$ (-2.812 eV).
Thus, the depth of the potential energy surface of 
the NO- and O$_2$-adsorbed Cu(OH)$_2$/$\gamma$-alumina 
are assumed to be nearly equal.

\section{\label{SEC5}Conclusions}

On the basis of the DFT calculations using the computational models for the solid surface,
we investigated
the regioselectivity of the Cu adsorption on the $\gamma$-alumina surface
and that of the NO adsorption on the Cu/$\gamma$-alumina surface
as well as the role of the $\gamma$-alumina support on the NO reduction.
The SSHT slab and cluster models of the $\gamma$-alumina surface,
where the reactive sites for the hydrogen termination are determined 
based on the crystal and molecular orbital coefficients,
reflect the experimental energy gap as well as the IR and UV/Vis spectra.
The obtained models have O, OH, and H$_2$O on the surfaces.
Notably,
the SSHT procedure contains an arbitrariness on deciding the sites for the hydrogen termination
when a few O atoms have orbital coefficients with similar magnitudes.
However, it is expected even on the real surface that 
the adsorption sites are not uniquely determined due to the amorphous structure.
Thus, this approach provides one way to determine 
the hydrogen termination sites on the metal oxide surfaces.
These sites can also be determined 
utilizing the VCD as a reactivity index,
although in the current study the orbital coefficients were used for simplicity.

The regioselectivity of the Cu adsorption on the $\gamma$-alumina surface
and that of the NO adsorption on the Cu/$\gamma$-alumina surface 
were elucidated by the VCD theory more clearly 
than by the frontier orbital theory
because of the localization of the vibrational states.
The anchor site for Cu(OH)$_2$ was determined to be 
the O atom located in the tetrahedral sites of the $\gamma$-alumina surface.
This O atom, which is not terminated with H atom,
has a lone pair and serves as the Lewis base.
The tetrahedral Al species transforms the structure upon the Cu loading
due to the Jahn--Teller distortion 
induced by the electron transfer from the Al species to Cu.
The VCD analyses demonstrate that
the NO adsorbs on the Cu(OH)$_2$,
and the second NO successively adsorbs on the initially adsorbed NO.
Namely, the dimerization of NO occurs on Cu(OH)$_2$.
The adsorption of NO on the Cu/$\gamma$-alumina easily occurs compared with O$_2$
because of the strongly bounded Cu--NO bond.

The role of the $\gamma$-alumina support is discovered
to raise the SOMO of the Cu catalyst,
which promotes the electron transfer from the Cu/$\gamma$-alumina to NO
because of the small orbital energy gap between them.
Consequently, 
the orbital of (NO)$_2$ with the N--N bonding and N--O anti-bonding
is occupied by electrons,
and this is considered to be the driving force of the NO reduction.
Another role of the $\gamma$-alumina support is 
to bond the Cu(OH)$_2$ on its surface to inhibit the catalyst aggregations.
In addition, the support acts as a heat bath 
for dissipating the excessive reaction energies due to the NO adsorption and reduction.
These roles of the catalyst support can be regarded as the anchoring effect.

\begin{acknowledgement}
This study was supported by 
Element Strategy Initiative of MEXT Grant Number JPMXP0112101003 and
JSPS KAKENHI Grant Number JP18K05261 in Scientific Research (C).
Numerical calculations were partly performed at
Supercomputer System, Institute for Chemical Research, Kyoto University,
Academic Center for Computing and Media Studies (ACCMS), Kyoto University,
and Research Center for Computational Science, Okazaki.
We would like to thank Editage (www.editage.com) for English language editing.
\end{acknowledgement}

\begin{suppinfo}
Band structure of the SSHT slab model;
Cartesian coordinates of the SSHT slab model;
molecular orbitals of the bare cluster;
orbital levels of the SSHT cluster model;
Cartesian coordinates of the SSHT cluster model;
OVCCs and OVCD of the $\gamma$-alumina cluster model;
geometry-optimized structures of the NO-adsorbed Cu(OH)$_2$/$\gamma$-alumina cluster model;
electron density differences of the Cu(OH)$_2$/$\gamma$-alumina cluster model;
geometry-optimized structures using the slab models;
atomic labels of the tetrahedral Al species;
interatomic distance and Mulliken charge of the tetrahedral Al species;
orbital levels of AlO$_4$;
VCCs of AlO$_4$;
geometry-optimized structures of the O$_2$-adsorbed Cu(OH)$_2$/$\gamma$-alumina;
fragment orbital analyses of the NO- and O$_2$-adsorbed Cu(OH)$_2$/$\gamma$-alumina.
\end{suppinfo}

\bibliography{refs}

\providecommand{\latin}[1]{#1}
\providecommand*\mcitethebibliography{\thebibliography}
\csname @ifundefined\endcsname{endmcitethebibliography}
  {\let\endmcitethebibliography\endthebibliography}{}
\begin{mcitethebibliography}{56}
\providecommand*\natexlab[1]{#1}
\providecommand*\mciteSetBstSublistMode[1]{}
\providecommand*\mciteSetBstMaxWidthForm[2]{}
\providecommand*\mciteBstWouldAddEndPuncttrue
  {\def\EndOfBibitem{\unskip.}}
\providecommand*\mciteBstWouldAddEndPunctfalse
  {\let\EndOfBibitem\relax}
\providecommand*\mciteSetBstMidEndSepPunct[3]{}
\providecommand*\mciteSetBstSublistLabelBeginEnd[3]{}
\providecommand*\EndOfBibitem{}
\mciteSetBstSublistMode{f}
\mciteSetBstMaxWidthForm{subitem}{(\alph{mcitesubitemcount})}
\mciteSetBstSublistLabelBeginEnd
  {\mcitemaxwidthsubitemform\space}
  {\relax}
  {\relax}

\bibitem[Ka{\v{s}}par \latin{et~al.}(2003)Ka{\v{s}}par, Fornasiero, and
  Hickey]{Kavspar2003_419}
Ka{\v{s}}par,~J.; Fornasiero,~P.; Hickey,~N. Automotive catalytic converters:
  current status and some perspectives. \emph{Catal. Today} \textbf{2003},
  \emph{77}, 419--449\relax
\mciteBstWouldAddEndPuncttrue
\mciteSetBstMidEndSepPunct{\mcitedefaultmidpunct}
{\mcitedefaultendpunct}{\mcitedefaultseppunct}\relax
\EndOfBibitem
\bibitem[Shelef and Graham(1994)Shelef, and Graham]{Shelef1994_433}
Shelef,~M.; Graham,~G. Why rhodium in automotive three-way catalysts?
  \emph{Catal. Rev.} \textbf{1994}, \emph{36}, 433--457\relax
\mciteBstWouldAddEndPuncttrue
\mciteSetBstMidEndSepPunct{\mcitedefaultmidpunct}
{\mcitedefaultendpunct}{\mcitedefaultseppunct}\relax
\EndOfBibitem
\bibitem[Brown and King(2000)Brown, and King]{Brown2000_2578}
Brown,~W.~A.; King,~D.~A. NO chemisorption and reactions on metal surfaces: a
  new perspective. \emph{J. Phys. Chem. B} \textbf{2000}, \emph{104},
  2578--2595\relax
\mciteBstWouldAddEndPuncttrue
\mciteSetBstMidEndSepPunct{\mcitedefaultmidpunct}
{\mcitedefaultendpunct}{\mcitedefaultseppunct}\relax
\EndOfBibitem
\bibitem[Deushi \latin{et~al.}(2017)Deushi, Ishikawa, and
  Nakai]{Deushi2017_15272}
Deushi,~F.; Ishikawa,~A.; Nakai,~H. Density Functional Theory Analysis of
  Elementary Reactions in NO$_x$ Reduction on Rh Surfaces and Rh Clusters.
  \emph{J. Phys. Chem. C} \textbf{2017}, \emph{121}, 15272--15281\relax
\mciteBstWouldAddEndPuncttrue
\mciteSetBstMidEndSepPunct{\mcitedefaultmidpunct}
{\mcitedefaultendpunct}{\mcitedefaultseppunct}\relax
\EndOfBibitem
\bibitem[Ishikawa and Tateyama(2018)Ishikawa, and Tateyama]{Ishikawa2018_17378}
Ishikawa,~A.; Tateyama,~Y. First-Principles Microkinetic Analysis of NO+CO
  Reactions on Rh (111) Surface toward Understanding NO$_x$ Reduction Pathways.
  \emph{J. Phys. Chem. C} \textbf{2018}, \emph{122}, 17378--17388\relax
\mciteBstWouldAddEndPuncttrue
\mciteSetBstMidEndSepPunct{\mcitedefaultmidpunct}
{\mcitedefaultendpunct}{\mcitedefaultseppunct}\relax
\EndOfBibitem
\bibitem[Takagi \latin{et~al.}(2019)Takagi, Ishimura, Fukuda, Ehara, and
  Sakaki]{Takagi2019_7021}
Takagi,~N.; Ishimura,~K.; Fukuda,~R.; Ehara,~M.; Sakaki,~S. Reaction Behavior
  of the NO Molecule on the Surface of an M n Particle (M= Ru, Rh, Pd, and Ag;
  $n$=13 and 55): Theoretical Study of Its Dependence on Transition-Metal
  Element. \emph{J. Phys. Chem. A} \textbf{2019}, \emph{123}, 7021--7033\relax
\mciteBstWouldAddEndPuncttrue
\mciteSetBstMidEndSepPunct{\mcitedefaultmidpunct}
{\mcitedefaultendpunct}{\mcitedefaultseppunct}\relax
\EndOfBibitem
\bibitem[Ward \latin{et~al.}(1993)Ward, Alemany, and Hoffmann]{Ward1993_7691}
Ward,~T.~R.; Alemany,~P.; Hoffmann,~R. Adhesion of rhodium, palladium, and
  platinum to alumina and the reduction of nitric oxide on the resulting
  surfaces: a theoretical analysis. \emph{J. Phys. Chem.} \textbf{1993},
  \emph{97}, 7691--7699\relax
\mciteBstWouldAddEndPuncttrue
\mciteSetBstMidEndSepPunct{\mcitedefaultmidpunct}
{\mcitedefaultendpunct}{\mcitedefaultseppunct}\relax
\EndOfBibitem
\bibitem[Ward \latin{et~al.}(1993)Ward, Hoffmann, and Shelef]{Ward1993_85}
Ward,~T.~R.; Hoffmann,~R.; Shelef,~M. Coupling nitrosyls as the first step in
  the reduction of NO on metal surfaces: the special role of rhodium.
  \emph{Surf. Sci.} \textbf{1993}, \emph{289}, 85--99\relax
\mciteBstWouldAddEndPuncttrue
\mciteSetBstMidEndSepPunct{\mcitedefaultmidpunct}
{\mcitedefaultendpunct}{\mcitedefaultseppunct}\relax
\EndOfBibitem
\bibitem[Yamamoto \latin{et~al.}(2002)Yamamoto, Tanaka, Kuma, Suzuki, Amano,
  Shimooka, Kohno, Funabiki, and Yoshida]{Yamamoto2002_2449}
Yamamoto,~T.; Tanaka,~T.; Kuma,~R.; Suzuki,~S.; Amano,~F.; Shimooka,~Y.;
  Kohno,~Y.; Funabiki,~T.; Yoshida,~S. NO reduction with CO in the presence of
  O$_2$ over Al$_2$O$_3$-supported and Cu-based catalysts. \emph{Phys. Chem.
  Chem. Phys.} \textbf{2002}, \emph{4}, 2449--2458\relax
\mciteBstWouldAddEndPuncttrue
\mciteSetBstMidEndSepPunct{\mcitedefaultmidpunct}
{\mcitedefaultendpunct}{\mcitedefaultseppunct}\relax
\EndOfBibitem
\bibitem[Yamamoto \latin{et~al.}(2002)Yamamoto, Tanaka, Suzuki, Kuma, Teramura,
  Kou, Funabiki, and Yoshida]{Yamamoto2002_113}
Yamamoto,~T.; Tanaka,~T.; Suzuki,~S.; Kuma,~R.; Teramura,~K.; Kou,~Y.;
  Funabiki,~T.; Yoshida,~S. NO reduction with CO in the presence of O$_2$ over
  Cu/Al$_2$O$_3$ (3)--structural analysis of active species by means of XAFS
  and UV/VIS/NIR spectroscopy. \emph{Top. Catal.} \textbf{2002}, \emph{18},
  113--118\relax
\mciteBstWouldAddEndPuncttrue
\mciteSetBstMidEndSepPunct{\mcitedefaultmidpunct}
{\mcitedefaultendpunct}{\mcitedefaultseppunct}\relax
\EndOfBibitem
\bibitem[Amano \latin{et~al.}(2006)Amano, Suzuki, Yamamoto, and
  Tanaka]{Amano2006_282}
Amano,~F.; Suzuki,~S.; Yamamoto,~T.; Tanaka,~T. One-electron reducibility of
  isolated copper oxide on alumina for selective NO--CO reaction. \emph{Appl.
  Catal. B: Environ.} \textbf{2006}, \emph{64}, 282--289\relax
\mciteBstWouldAddEndPuncttrue
\mciteSetBstMidEndSepPunct{\mcitedefaultmidpunct}
{\mcitedefaultendpunct}{\mcitedefaultseppunct}\relax
\EndOfBibitem
\bibitem[Hosokawa \latin{et~al.}(2017)Hosokawa, Matsuki, Tamaru, Oshino,
  Aritani, Asakura, Teramura, and Tanaka]{Hosokawa2017_74}
Hosokawa,~S.; Matsuki,~K.; Tamaru,~K.; Oshino,~Y.; Aritani,~H.; Asakura,~H.;
  Teramura,~K.; Tanaka,~T. Selective reduction of NO over Cu/Al$_2$O$_3$:
  Enhanced catalytic activity by infinitesimal loading of Rh on Cu/Al$_2$O$_3$.
  \emph{Molecular Catalysis} \textbf{2017}, \emph{442}, 74--82\relax
\mciteBstWouldAddEndPuncttrue
\mciteSetBstMidEndSepPunct{\mcitedefaultmidpunct}
{\mcitedefaultendpunct}{\mcitedefaultseppunct}\relax
\EndOfBibitem
\bibitem[Fukuda \latin{et~al.}(2018)Fukuda, Sakai, Takagi, Matsui, Ehara,
  Hosokawa, Tanaka, and Sakaki]{Fukuda2018_3833}
Fukuda,~R.; Sakai,~S.; Takagi,~N.; Matsui,~M.; Ehara,~M.; Hosokawa,~S.;
  Tanaka,~T.; Sakaki,~S. Mechanism of NO--CO reaction over highly dispersed
  cuprous oxide on $\gamma$-alumina catalyst using a metal--support interfacial
  site in the presence of oxygen: similarities to and differences from
  biological systems. \emph{Catal. Sci. Technol.} \textbf{2018}, \emph{8},
  3833--3845\relax
\mciteBstWouldAddEndPuncttrue
\mciteSetBstMidEndSepPunct{\mcitedefaultmidpunct}
{\mcitedefaultendpunct}{\mcitedefaultseppunct}\relax
\EndOfBibitem
\bibitem[Nagai \latin{et~al.}(2006)Nagai, Hirabayashi, Dohmae, Takagi, Minami,
  Shinjoh, and Matsumoto]{Nagai2006_103}
Nagai,~Y.; Hirabayashi,~T.; Dohmae,~K.; Takagi,~N.; Minami,~T.; Shinjoh,~H.;
  Matsumoto,~S. Sintering inhibition mechanism of platinum supported on
  ceria-based oxide and Pt-oxide--support interaction. \emph{J. Catal.}
  \textbf{2006}, \emph{242}, 103--109\relax
\mciteBstWouldAddEndPuncttrue
\mciteSetBstMidEndSepPunct{\mcitedefaultmidpunct}
{\mcitedefaultendpunct}{\mcitedefaultseppunct}\relax
\EndOfBibitem
\bibitem[Machida \latin{et~al.}(2014)Machida, Minami, Ikeue, Hinokuma, Nagao,
  Sato, and Nakahara]{Machida2014_5799}
Machida,~M.; Minami,~S.; Ikeue,~K.; Hinokuma,~S.; Nagao,~Y.; Sato,~T.;
  Nakahara,~Y. Rhodium nanoparticle anchoring on AlPO$_4$ for efficient
  catalyst sintering suppression. \emph{Chem. Mater.} \textbf{2014}, \emph{26},
  5799--5805\relax
\mciteBstWouldAddEndPuncttrue
\mciteSetBstMidEndSepPunct{\mcitedefaultmidpunct}
{\mcitedefaultendpunct}{\mcitedefaultseppunct}\relax
\EndOfBibitem
\bibitem[Trueba and Trasatti(2005)Trueba, and Trasatti]{Trueba2005_3393}
Trueba,~M.; Trasatti,~S.~P. $\gamma$-alumina as a support for catalysts: a
  review of fundamental aspects. \emph{Eur. J. Inorg. Chem.} \textbf{2005},
  \emph{17}, 3393--3403\relax
\mciteBstWouldAddEndPuncttrue
\mciteSetBstMidEndSepPunct{\mcitedefaultmidpunct}
{\mcitedefaultendpunct}{\mcitedefaultseppunct}\relax
\EndOfBibitem
\bibitem[Fukui \latin{et~al.}(1952)Fukui, Yonezawa, and Shingu]{Fukui1952_722}
Fukui,~K.; Yonezawa,~T.; Shingu,~H. A molecular orbital theory of reactivity in
  aromatic hydrocarbons. \emph{J. Chem. Phys.} \textbf{1952}, \emph{20},
  722--725\relax
\mciteBstWouldAddEndPuncttrue
\mciteSetBstMidEndSepPunct{\mcitedefaultmidpunct}
{\mcitedefaultendpunct}{\mcitedefaultseppunct}\relax
\EndOfBibitem
\bibitem[Fukui(1982)]{Fukui1982_747}
Fukui,~K. Role of frontier orbitals in chemical reactions. \emph{Science}
  \textbf{1982}, \emph{218}, 747--754\relax
\mciteBstWouldAddEndPuncttrue
\mciteSetBstMidEndSepPunct{\mcitedefaultmidpunct}
{\mcitedefaultendpunct}{\mcitedefaultseppunct}\relax
\EndOfBibitem
\bibitem[Sato \latin{et~al.}(2008)Sato, Tokunaga, and Tanaka]{Sato2008_758}
Sato,~T.; Tokunaga,~K.; Tanaka,~K. Vibronic coupling in naphthalene anion:
  vibronic coupling density analysis for totally symmetric vibrational modes.
  \emph{J. Phys. Chem. A} \textbf{2008}, \emph{112}, 758--767\relax
\mciteBstWouldAddEndPuncttrue
\mciteSetBstMidEndSepPunct{\mcitedefaultmidpunct}
{\mcitedefaultendpunct}{\mcitedefaultseppunct}\relax
\EndOfBibitem
\bibitem[Kojima \latin{et~al.}(2019)Kojima, Ota, Teramura, Hosokawa, Tanaka,
  and Sato]{Kojima2019_239}
Kojima,~Y.; Ota,~W.; Teramura,~K.; Hosokawa,~S.; Tanaka,~T.; Sato,~T. Model
  building of metal oxide surfaces and vibronic coupling density as a
  reactivity index: Regioselectivity of CO$_2$ adsorption on Ag-loaded
  Ga$_2$O$_3$. \emph{Chem. Phys. Lett.} \textbf{2019}, \emph{715},
  239--243\relax
\mciteBstWouldAddEndPuncttrue
\mciteSetBstMidEndSepPunct{\mcitedefaultmidpunct}
{\mcitedefaultendpunct}{\mcitedefaultseppunct}\relax
\EndOfBibitem
\bibitem[Ota \latin{et~al.}(2018)Ota, Teramura, Hosokawa, Tanaka, and
  Sato]{Ota2018_138}
Ota,~W.; Teramura,~K.; Hosokawa,~S.; Tanaka,~T.; Sato,~T. Regioselectivity of
  H$_2$ adsorption on Ga$_2$O$_3$ surface based on vibronic coupling density
  analysis. \emph{J. Comput. Chem. Jpn.} \textbf{2018}, \emph{17},
  138--141\relax
\mciteBstWouldAddEndPuncttrue
\mciteSetBstMidEndSepPunct{\mcitedefaultmidpunct}
{\mcitedefaultendpunct}{\mcitedefaultseppunct}\relax
\EndOfBibitem
\bibitem[Sato \latin{et~al.}(2012)Sato, Iwahara, Haruta, and
  Tanaka]{Sato2012_257}
Sato,~T.; Iwahara,~N.; Haruta,~N.; Tanaka,~K. C$_{60}$ bearing ethylene
  moieties. \emph{Chem. Phys. Lett.} \textbf{2012}, \emph{531}, 257--260\relax
\mciteBstWouldAddEndPuncttrue
\mciteSetBstMidEndSepPunct{\mcitedefaultmidpunct}
{\mcitedefaultendpunct}{\mcitedefaultseppunct}\relax
\EndOfBibitem
\bibitem[Haruta \latin{et~al.}(2012)Haruta, Sato, and Tanaka]{Haruta2012_9702}
Haruta,~N.; Sato,~T.; Tanaka,~K. Chemical reactivity in nucleophilic
  cycloaddition to C$_{70}$: vibronic coupling density and vibronic coupling
  constants as reactivity indices. \emph{J. Org. Chem.} \textbf{2012},
  \emph{77}, 9702--9706\relax
\mciteBstWouldAddEndPuncttrue
\mciteSetBstMidEndSepPunct{\mcitedefaultmidpunct}
{\mcitedefaultendpunct}{\mcitedefaultseppunct}\relax
\EndOfBibitem
\bibitem[Haruta \latin{et~al.}(2013)Haruta, Sato, Iwahara, and
  Tanaka]{Haruta2013_012003}
Haruta,~N.; Sato,~T.; Iwahara,~N.; Tanaka,~K. Vibronic couplings in
  cycloadditions to fullerenes. \emph{J. Phys.: Conf. Ser.} \textbf{2013},
  \emph{428}, 012003\relax
\mciteBstWouldAddEndPuncttrue
\mciteSetBstMidEndSepPunct{\mcitedefaultmidpunct}
{\mcitedefaultendpunct}{\mcitedefaultseppunct}\relax
\EndOfBibitem
\bibitem[Haruta \latin{et~al.}(2014)Haruta, Sato, and Tanaka]{Haruta2014_3510}
Haruta,~N.; Sato,~T.; Tanaka,~K. Regioselectivity in multiple cycloadditions to
  fullerene C$_{60}$: vibronic coupling density analysis. \emph{Tetrahedron}
  \textbf{2014}, \emph{70}, 3510--3513\relax
\mciteBstWouldAddEndPuncttrue
\mciteSetBstMidEndSepPunct{\mcitedefaultmidpunct}
{\mcitedefaultendpunct}{\mcitedefaultseppunct}\relax
\EndOfBibitem
\bibitem[Haruta \latin{et~al.}(2014)Haruta, Sato, and Tanaka]{Haruta2014_141}
Haruta,~N.; Sato,~T.; Tanaka,~K. Reactivity of endohedral metallofullerene
  La$_{2}$@C$_{80}$ in nucleophilic and electrophilic attacks: vibronic
  coupling density approach. \emph{J. Org. Chem.} \textbf{2014}, \emph{80},
  141--147\relax
\mciteBstWouldAddEndPuncttrue
\mciteSetBstMidEndSepPunct{\mcitedefaultmidpunct}
{\mcitedefaultendpunct}{\mcitedefaultseppunct}\relax
\EndOfBibitem
\bibitem[Haruta \latin{et~al.}(2015)Haruta, Sato, and Tanaka]{Haruta2015_590}
Haruta,~N.; Sato,~T.; Tanaka,~K. Reactivity index for Diels--Alder
  cycloadditions to large polycyclic aromatic hydrocarbons using vibronic
  coupling density. \emph{Tetrahedron Lett.} \textbf{2015}, \emph{56},
  590--594\relax
\mciteBstWouldAddEndPuncttrue
\mciteSetBstMidEndSepPunct{\mcitedefaultmidpunct}
{\mcitedefaultendpunct}{\mcitedefaultseppunct}\relax
\EndOfBibitem
\bibitem[Hoffman(1988)]{Hoffman1988}
Hoffman,~R. \emph{Solids and Surfaces: A Chemist's View of Bonding in Extended
  Structures}; VCH: New York, 1988\relax
\mciteBstWouldAddEndPuncttrue
\mciteSetBstMidEndSepPunct{\mcitedefaultmidpunct}
{\mcitedefaultendpunct}{\mcitedefaultseppunct}\relax
\EndOfBibitem
\bibitem[Sato \latin{et~al.}(2009)Sato, Tokunaga, Iwahara, Shizu, and
  Tanaka]{Sato2009_99}
Sato,~T.; Tokunaga,~K.; Iwahara,~N.; Shizu,~K.; Tanaka,~K. Vibronic Coupling
  Constant and Vibronic Coupling Density. In \emph{The Jahn-Teller Effect:
  Fundamentals and Implications for Physics and Chemistry}; K{\"o}ppel,~H.,
  Yarkony,~D.~R., Barentzen,~H., Eds.; Springer-Verlag: Berlin and Hidelberg,
  2009; pp 99--129\relax
\mciteBstWouldAddEndPuncttrue
\mciteSetBstMidEndSepPunct{\mcitedefaultmidpunct}
{\mcitedefaultendpunct}{\mcitedefaultseppunct}\relax
\EndOfBibitem
\bibitem[Sato \latin{et~al.}(2013)Sato, Uejima, Iwahara, Haruta, Shizu, and
  Tanaka]{Sato2013_012010}
Sato,~T.; Uejima,~M.; Iwahara,~N.; Haruta,~N.; Shizu,~K.; Tanaka,~K. Vibronic
  coupling density and related concepts. \emph{J. Phys.: Conf. Ser.}
  \textbf{2013}, \emph{428}, 012010\relax
\mciteBstWouldAddEndPuncttrue
\mciteSetBstMidEndSepPunct{\mcitedefaultmidpunct}
{\mcitedefaultendpunct}{\mcitedefaultseppunct}\relax
\EndOfBibitem
\bibitem[Fischer(1984)]{Fischer1984}
Fischer,~G. \emph{Vibronic Coupling: The Interaction between the Electronic and
  Nuclear Motions}; Academic Press: London, 1984\relax
\mciteBstWouldAddEndPuncttrue
\mciteSetBstMidEndSepPunct{\mcitedefaultmidpunct}
{\mcitedefaultendpunct}{\mcitedefaultseppunct}\relax
\EndOfBibitem
\bibitem[Azumi and Matsuzaki(1977)Azumi, and Matsuzaki]{Azumi1977_315}
Azumi,~T.; Matsuzaki,~K. What Does the Term "Vibronic Coupling" Mean?
  \emph{Photochem. Photobiol.} \textbf{1977}, \emph{25}, 315--326\relax
\mciteBstWouldAddEndPuncttrue
\mciteSetBstMidEndSepPunct{\mcitedefaultmidpunct}
{\mcitedefaultendpunct}{\mcitedefaultseppunct}\relax
\EndOfBibitem
\bibitem[Hellmann(1937)]{Hellmann1937}
Hellmann,~H. \emph{Einf\"uhrung in die Quantenchemie}; Deuticke and Company:
  Leipzog, 1937\relax
\mciteBstWouldAddEndPuncttrue
\mciteSetBstMidEndSepPunct{\mcitedefaultmidpunct}
{\mcitedefaultendpunct}{\mcitedefaultseppunct}\relax
\EndOfBibitem
\bibitem[Feynman(1939)]{Feynman1939_340}
Feynman,~R.~P. Forces in molecules. \emph{Phys. Rev.} \textbf{1939}, \emph{56},
  340--343\relax
\mciteBstWouldAddEndPuncttrue
\mciteSetBstMidEndSepPunct{\mcitedefaultmidpunct}
{\mcitedefaultendpunct}{\mcitedefaultseppunct}\relax
\EndOfBibitem
\bibitem[Parr and Yang(1984)Parr, and Yang]{Parr1984_4049}
Parr,~R.~G.; Yang,~W. Density functional approach to the frontier-electron
  theory of chemical reactivity. \emph{J. Am. Chem. Soc.} \textbf{1984},
  \emph{106}, 4049--4050\relax
\mciteBstWouldAddEndPuncttrue
\mciteSetBstMidEndSepPunct{\mcitedefaultmidpunct}
{\mcitedefaultendpunct}{\mcitedefaultseppunct}\relax
\EndOfBibitem
\bibitem[Parr and Yang(1994)Parr, and Yang]{Parr1994}
Parr,~R.~G.; Yang,~W. \emph{Density-Functional Theory of Atoms and Molecules};
  Oxford University Press: New York, 1994\relax
\mciteBstWouldAddEndPuncttrue
\mciteSetBstMidEndSepPunct{\mcitedefaultmidpunct}
{\mcitedefaultendpunct}{\mcitedefaultseppunct}\relax
\EndOfBibitem
\bibitem[Hoffmann \latin{et~al.}(1973)Hoffmann, Fujimoto, Swenson, and
  Wan]{Hoffmann1973_7644}
Hoffmann,~R.; Fujimoto,~H.; Swenson,~J.~R.; Wan,~C.-C. Theoretical aspects of
  the bonding in some three-membered rings containing sulfur. \emph{J. Am.
  Chem. Soc.} \textbf{1973}, \emph{95}, 7644--7650\relax
\mciteBstWouldAddEndPuncttrue
\mciteSetBstMidEndSepPunct{\mcitedefaultmidpunct}
{\mcitedefaultendpunct}{\mcitedefaultseppunct}\relax
\EndOfBibitem
\bibitem[Krokidis \latin{et~al.}(2001)Krokidis, Raybaud, Gobichon, Rebours,
  Euzen, and Toulhoat]{Krokidis2001_5121}
Krokidis,~X.; Raybaud,~P.; Gobichon,~A.-E.; Rebours,~B.; Euzen,~P.;
  Toulhoat,~H. Theoretical study of the dehydration process of boehmite to
  $\gamma$-alumina. \emph{J. Phys. Chem. B} \textbf{2001}, \emph{105},
  5121--5130\relax
\mciteBstWouldAddEndPuncttrue
\mciteSetBstMidEndSepPunct{\mcitedefaultmidpunct}
{\mcitedefaultendpunct}{\mcitedefaultseppunct}\relax
\EndOfBibitem
\bibitem[Digne \latin{et~al.}(2002)Digne, Sautet, Raybaud, Euzen, and
  Toulhoat]{Digne2002_1}
Digne,~M.; Sautet,~P.; Raybaud,~P.; Euzen,~P.; Toulhoat,~H. Hydroxyl groups on
  $\gamma$-alumina surfaces: A DFT study. \emph{J. Catal.} \textbf{2002},
  \emph{211}, 1--5\relax
\mciteBstWouldAddEndPuncttrue
\mciteSetBstMidEndSepPunct{\mcitedefaultmidpunct}
{\mcitedefaultendpunct}{\mcitedefaultseppunct}\relax
\EndOfBibitem
\bibitem[Digne \latin{et~al.}(2004)Digne, Sautet, Raybaud, Euzen, and
  Toulhoat]{Digne2004_54}
Digne,~M.; Sautet,~P.; Raybaud,~P.; Euzen,~P.; Toulhoat,~H. Use of DFT to
  achieve a rational understanding of acid--basic properties of
  $\gamma$-alumina surfaces. \emph{J. Catal.} \textbf{2004}, \emph{226},
  54--68\relax
\mciteBstWouldAddEndPuncttrue
\mciteSetBstMidEndSepPunct{\mcitedefaultmidpunct}
{\mcitedefaultendpunct}{\mcitedefaultseppunct}\relax
\EndOfBibitem
\bibitem[Nortier \latin{et~al.}(1990)Nortier, Fourre, Saad, Saur, and
  Lavalley]{Nortier1990_141}
Nortier,~P.; Fourre,~P.; Saad,~A.~M.; Saur,~O.; Lavalley,~J. Effects of
  crystallinity and morphology on the surface properties ofalumina. \emph{App.
  Catal.} \textbf{1990}, \emph{61}, 141--160\relax
\mciteBstWouldAddEndPuncttrue
\mciteSetBstMidEndSepPunct{\mcitedefaultmidpunct}
{\mcitedefaultendpunct}{\mcitedefaultseppunct}\relax
\EndOfBibitem
\bibitem[Morterra and Magnacca(1996)Morterra, and Magnacca]{Morterra1996_497}
Morterra,~C.; Magnacca,~G. A case study: surface chemistry and surface
  structure of catalytic aluminas, as studied by vibrational spectroscopy of
  adsorbed species. \emph{Catal. Today} \textbf{1996}, \emph{27},
  497--532\relax
\mciteBstWouldAddEndPuncttrue
\mciteSetBstMidEndSepPunct{\mcitedefaultmidpunct}
{\mcitedefaultendpunct}{\mcitedefaultseppunct}\relax
\EndOfBibitem
\bibitem[Busca \latin{et~al.}(1993)Busca, Lorenzelli, Ramis, and
  Willey]{Busca1993_1492}
Busca,~G.; Lorenzelli,~V.; Ramis,~G.; Willey,~R.~J. Surface sites on
  spinel-type and corundum-type metal oxide powders. \emph{Langmuir}
  \textbf{1993}, \emph{9}, 1492--1499\relax
\mciteBstWouldAddEndPuncttrue
\mciteSetBstMidEndSepPunct{\mcitedefaultmidpunct}
{\mcitedefaultendpunct}{\mcitedefaultseppunct}\relax
\EndOfBibitem
\bibitem[Tsyganenko and Mardilovich(1996)Tsyganenko, and
  Mardilovich]{Tsyganenko1996_4843}
Tsyganenko,~A.~A.; Mardilovich,~P.~P. Structure of alumina surfaces. \emph{J.
  Chem. Soc., Faraday Trans.} \textbf{1996}, \emph{92}, 4843--4852\relax
\mciteBstWouldAddEndPuncttrue
\mciteSetBstMidEndSepPunct{\mcitedefaultmidpunct}
{\mcitedefaultendpunct}{\mcitedefaultseppunct}\relax
\EndOfBibitem
\bibitem[Pecharroman \latin{et~al.}(1999)Pecharroman, Sobrados, Iglesias,
  Gonzalez-Carreno, and Sanz]{Pecharroman1999_6160}
Pecharroman,~C.; Sobrados,~I.; Iglesias,~J.; Gonzalez-Carreno,~T.; Sanz,~J.
  Thermal evolution of transitional aluminas followed by NMR and IR
  spectroscopies. \emph{J. Phys. Chem. B} \textbf{1999}, \emph{103},
  6160--6170\relax
\mciteBstWouldAddEndPuncttrue
\mciteSetBstMidEndSepPunct{\mcitedefaultmidpunct}
{\mcitedefaultendpunct}{\mcitedefaultseppunct}\relax
\EndOfBibitem
\bibitem[Paglia \latin{et~al.}(2004)Paglia, Buckley, Udovic, Rohl, Jones,
  Maitland, and Connolly]{Paglia2004_1914}
Paglia,~G.; Buckley,~C.~E.; Udovic,~T.~J.; Rohl,~A.~L.; Jones,~F.;
  Maitland,~C.~F.; Connolly,~J. Boehmite-derived $\gamma$-alumina system. 2.
  Consideration of hydrogen and surface effects. \emph{Chem. Mater.}
  \textbf{2004}, \emph{16}, 1914--1923\relax
\mciteBstWouldAddEndPuncttrue
\mciteSetBstMidEndSepPunct{\mcitedefaultmidpunct}
{\mcitedefaultendpunct}{\mcitedefaultseppunct}\relax
\EndOfBibitem
\bibitem[Frisch \latin{et~al.}(2013)Frisch, Trucks, Schlegel, Scuseria, Robb,
  Cheeseman, Scalmani, Barone, Mennucci, Petersson, Nakatsuji, Caricato, Li,
  Hratchian, Izmaylov, Bloino, Zheng, Sonnenberg, Hada, Ehara, Toyota, Fukuda,
  Hasegawa, Ishida, Nakajima, Honda, Kitao, Nakai, Vreven, Montgomery~Jr.,
  Peralta, Ogliaro, Bearpark, J, Brothers, Kudin, Staroverov, Keith, Kobayashi,
  Normand, Raghavachari, Rendell, Burant, Iyengar, Tomasi, Cossi, Rega, Millam,
  Klene, Knox, Cross, Bakken, Adamo, Jaramillo, Gomperts, Stratmann, Yazyev,
  Austin, Cammi, Pomelli, Ochterski, Martin, Morokuma, Zakrzewski, Voth,
  Salvador, Dannenberg, Dapprich, Daniels, Farkas, Foresman, Ortiz, Cioslowski,
  and Fox]{Frisch2013D}
Frisch,~M.~J.; Trucks,~G.~W.; Schlegel,~H.~B.; Scuseria,~G.~E.; Robb,~M.~A.;
  Cheeseman,~J.~R.; Scalmani,~G.; Barone,~V.; Mennucci,~B.; Petersson,~G.~A.;
  Nakatsuji,~H.; Caricato,~M.; Li,~X.; Hratchian,~H.~P.; Izmaylov,~A.~F.;
  Bloino,~J.; Zheng,~G.; Sonnenberg,~J.~L.; Hada,~M.; Ehara,~M.; Toyota,~K.;
  Fukuda,~R.; Hasegawa,~J.; Ishida,~M.; Nakajima,~T.; Honda,~Y.; Kitao,~O.;
  Nakai,~H.; Vreven,~T.; Montgomery~Jr.,~J.~A.; Peralta,~J.~E.; Ogliaro,~F.;
  Bearpark,~B.; J,~H.~J.; Brothers,~E.; Kudin,~K.~N.; Staroverov,~V.~N.;
  Keith,~T.; Kobayashi,~R.; Normand,~J.; Raghavachari,~K.; Rendell,~A.;
  Burant,~J.~C.; Iyengar,~S.~S.; Tomasi,~J.; Cossi,~M.; Rega,~N.;
  Millam,~J.~M.; Klene,~M.; Knox,~J.~E.; Cross,~J.~B.; Bakken,~V.; Adamo,~C.;
  Jaramillo,~J.; Gomperts,~R.; Stratmann,~R.~E.; Yazyev,~O.; Austin,~A.~J.;
  Cammi,~R.; Pomelli,~C.; Ochterski,~J.~W.; Martin,~R.~L.; Morokuma,~K.;
  Zakrzewski,~V.~G.; Voth,~G.~A.; Salvador,~P.; Dannenberg,~J.~J.;
  Dapprich,~S.; Daniels,~A.~D.; Farkas,~O.; Foresman,~J.~B.; Ortiz,~J.~V.;
  Cioslowski,~J.; Fox,~D.~J. Gaussian 09, Revision D. 01, Gaussian, Inc.:
  Wallingford, CT, 2013\relax
\mciteBstWouldAddEndPuncttrue
\mciteSetBstMidEndSepPunct{\mcitedefaultmidpunct}
{\mcitedefaultendpunct}{\mcitedefaultseppunct}\relax
\EndOfBibitem
\bibitem[Frisch \latin{et~al.}(2013)Frisch, Trucks, Schlegel, Scuseria, Robb,
  Cheeseman, Scalmani, Barone, Mennucci, Petersson, Nakatsuji, Caricato, Li,
  Hratchian, Izmaylov, Bloino, Zheng, Sonnenberg, Hada, Ehara, Toyota, Fukuda,
  Hasegawa, Ishida, Nakajima, Honda, Kitao, Nakai, Vreven, Montgomery~Jr.,
  Peralta, Ogliaro, Bearpark, J, Brothers, Kudin, Staroverov, Keith, Kobayashi,
  Normand, Raghavachari, Rendell, Burant, Iyengar, Tomasi, Cossi, Rega, Millam,
  Klene, Knox, Cross, Bakken, Adamo, Jaramillo, Gomperts, Stratmann, Yazyev,
  Austin, Cammi, Pomelli, Ochterski, Martin, Morokuma, Zakrzewski, Voth,
  Salvador, Dannenberg, Dapprich, Daniels, Farkas, Foresman, Ortiz, Cioslowski,
  and Fox]{Frisch2013E}
Frisch,~M.~J.; Trucks,~G.~W.; Schlegel,~H.~B.; Scuseria,~G.~E.; Robb,~M.~A.;
  Cheeseman,~J.~R.; Scalmani,~G.; Barone,~V.; Mennucci,~B.; Petersson,~G.~A.;
  Nakatsuji,~H.; Caricato,~M.; Li,~X.; Hratchian,~H.~P.; Izmaylov,~A.~F.;
  Bloino,~J.; Zheng,~G.; Sonnenberg,~J.~L.; Hada,~M.; Ehara,~M.; Toyota,~K.;
  Fukuda,~R.; Hasegawa,~J.; Ishida,~M.; Nakajima,~T.; Honda,~Y.; Kitao,~O.;
  Nakai,~H.; Vreven,~T.; Montgomery~Jr.,~J.~A.; Peralta,~J.~E.; Ogliaro,~F.;
  Bearpark,~B.; J,~H.~J.; Brothers,~E.; Kudin,~K.~N.; Staroverov,~V.~N.;
  Keith,~T.; Kobayashi,~R.; Normand,~J.; Raghavachari,~K.; Rendell,~A.;
  Burant,~J.~C.; Iyengar,~S.~S.; Tomasi,~J.; Cossi,~M.; Rega,~N.;
  Millam,~J.~M.; Klene,~M.; Knox,~J.~E.; Cross,~J.~B.; Bakken,~V.; Adamo,~C.;
  Jaramillo,~J.; Gomperts,~R.; Stratmann,~R.~E.; Yazyev,~O.; Austin,~A.~J.;
  Cammi,~R.; Pomelli,~C.; Ochterski,~J.~W.; Martin,~R.~L.; Morokuma,~K.;
  Zakrzewski,~V.~G.; Voth,~G.~A.; Salvador,~P.; Dannenberg,~J.~J.;
  Dapprich,~S.; Daniels,~A.~D.; Farkas,~O.; Foresman,~J.~B.; Ortiz,~J.~V.;
  Cioslowski,~J.; Fox,~D.~J. Gaussian 09, Revision E. 01, Gaussian, Inc.:
  Wallingford, CT, 2013\relax
\mciteBstWouldAddEndPuncttrue
\mciteSetBstMidEndSepPunct{\mcitedefaultmidpunct}
{\mcitedefaultendpunct}{\mcitedefaultseppunct}\relax
\EndOfBibitem
\bibitem[Hoffmann(1963)]{Hoffmann1963_1397}
Hoffmann,~R. An extended H{\"u}ckel theory. I. hydrocarbons. \emph{J. Chem.
  Phys.} \textbf{1963}, \emph{39}, 1397--1412\relax
\mciteBstWouldAddEndPuncttrue
\mciteSetBstMidEndSepPunct{\mcitedefaultmidpunct}
{\mcitedefaultendpunct}{\mcitedefaultseppunct}\relax
\EndOfBibitem
\bibitem[Landrum and Glassey()Landrum, and Glassey]{yaehmop}
Landrum,~G.; Glassey,~W. Yet Another Extended H{\"u}ckel Molecular Orbital
  Package (YAeHMOP). \url{http://yaehmop.sourceforge.net.}\relax
\mciteBstWouldAddEndPunctfalse
\mciteSetBstMidEndSepPunct{\mcitedefaultmidpunct}
{}{\mcitedefaultseppunct}\relax
\EndOfBibitem
\bibitem[Te~Velde and Baerends(1991)Te~Velde, and Baerends]{Velde1991_7888}
Te~Velde,~G.; Baerends,~E. Precise density-functional method for periodic
  structures. \emph{Phys. Rev. B} \textbf{1991}, \emph{44}, 7888\relax
\mciteBstWouldAddEndPuncttrue
\mciteSetBstMidEndSepPunct{\mcitedefaultmidpunct}
{\mcitedefaultendpunct}{\mcitedefaultseppunct}\relax
\EndOfBibitem
\bibitem[Te~Velde and Baerends(1992)Te~Velde, and Baerends]{Velde1992_84}
Te~Velde,~G.; Baerends,~E. Numerical integration for polyatomic systems.
  \emph{J. Comput. Phys.} \textbf{1992}, \emph{99}, 84--98\relax
\mciteBstWouldAddEndPuncttrue
\mciteSetBstMidEndSepPunct{\mcitedefaultmidpunct}
{\mcitedefaultendpunct}{\mcitedefaultseppunct}\relax
\EndOfBibitem
\bibitem[Ealet \latin{et~al.}(1994)Ealet, Elyakhloufi, Gillet, and
  Ricci]{Ealet1994_92}
Ealet,~B.; Elyakhloufi,~M.~H.; Gillet,~E.; Ricci,~M. Electronic and
  crystallographic structure of $\gamma$-alumina thin films. \emph{Thin Solid
  Films} \textbf{1994}, \emph{250}, 92--100\relax
\mciteBstWouldAddEndPuncttrue
\mciteSetBstMidEndSepPunct{\mcitedefaultmidpunct}
{\mcitedefaultendpunct}{\mcitedefaultseppunct}\relax
\EndOfBibitem
\bibitem[Bersuker(2006)]{Bersuker2006}
Bersuker,~I.~B. \emph{The Jahn--Teller Effect}; Cambridge University Press,
  2006\relax
\mciteBstWouldAddEndPuncttrue
\mciteSetBstMidEndSepPunct{\mcitedefaultmidpunct}
{\mcitedefaultendpunct}{\mcitedefaultseppunct}\relax
\EndOfBibitem
\bibitem[Ceulemans and Vanquickenborne(1989)Ceulemans, and
  Vanquickenborne]{Ceulemans1989_125}
Ceulemans,~A.; Vanquickenborne,~L. The epikernel principle. \emph{Struct.
  Bonding} \textbf{1989}, \emph{71}, 125--159\relax
\mciteBstWouldAddEndPuncttrue
\mciteSetBstMidEndSepPunct{\mcitedefaultmidpunct}
{\mcitedefaultendpunct}{\mcitedefaultseppunct}\relax
\EndOfBibitem
\end{mcitethebibliography}

\end{document}


\renewcommand{\thefigure}{S\arabic{figure}}
\renewcommand{\thetable}{S\arabic{table}}
\renewcommand{\thesection}{S\arabic{section}}

\vspace*{\stretch{1}}

\begin{flushleft}
{\LARGE \bf
Supporting Information
}\\
\end{flushleft}

\vspace*{\stretch{1}}

\begin{flushleft}
{\LARGE \bf
Role of Catalyst Support and
Regioselectivity of Molecular Adsorption 
on a Metal Oxide Surface: \\
NO Reduction on Cu/$\gamma$-alumina\\
} 
~\\
~\\
~\\
~\\
{\Large
Wataru Ota$^{1,2}$, 
Yasuro Kojima$^2$, 
Saburo Hosokawa$^{2,3}$,
Kentaro Teramura$^{2,3}$, 
Tsunehiro Tanaka$^{2,3}$, 
Tohru Sato$^{1,2,3}$*
}
~\\
~\\
~\\
{\Large
\textit{
$^1$ Fukui Institute for Fundamental Chemistry, Kyoto University, Sakyo-ku, Kyoto 606-8103, Japan
~\\
~\\
$^2$ Department of Molecular Engineering, Graduate School of Engineering, Kyoto University, Nishikyo-ku, Kyoto 615-8510, Japan
~\\
~\\
$^3$ 
Elements Strategy Initiative for Catalysts \& Batteries (ESICB), Kyoto University, Kyotodaigaku Katsura, Nishikyo-ku, Kyoto 615-8245, Japan
}
}
\end{flushleft}

\vspace*{\stretch{1}}

\newpage

\begin{figure}[!ht]
\begin{minipage}[c][1\textheight]{1\textwidth}
\centering
\includegraphics[width=0.6\hsize]{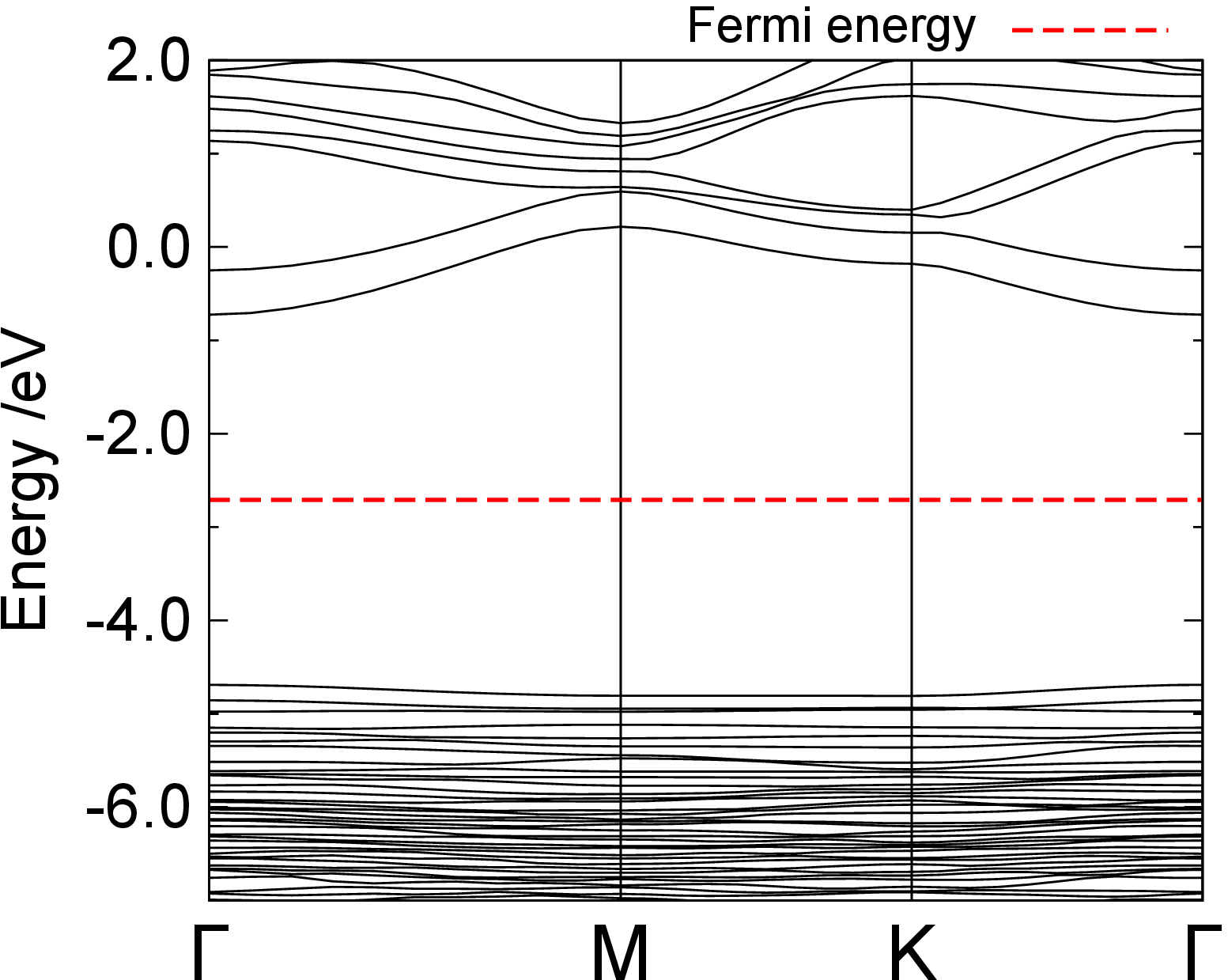}
\caption{
Calculated band structure of the SSHT slab model.
The band gap is 3.97 eV.
}
\label{FIGS1}
\end{minipage}
\end{figure}

\clearpage

\begin{table}[!h]
\begin{minipage}[c][1\textheight]{1\textwidth}
\caption{\label{TABLES1} 
Cartesian coordinates of the SSHT slab model.
Lattice vectors are 
(-8.0678, 0.0000, 0.0000) and 
(0.0000, 8.4130, 0.0000).
The coordinates are provided in Angstrom units.
}
\begin{center}
	\begin{tabular}{ccrrrccrrr} \hline\hline
No. & Atom & 
\multicolumn{1}{c}{$x$} & \multicolumn{1}{c}{$y$} & \multicolumn{1}{c}{$z$} &
No. & Atom & 
\multicolumn{1}{c}{$x$} & \multicolumn{1}{c}{$y$} & \multicolumn{1}{c}{$z$} \\ \hline
1  &   Al  &    -0.8532  &   -2.1445 &     -0.5612 &
33 &   O   &     1.3434   &   3.4186  &    -0.9256 \\ 
2  &   Al  &     1.1183   &   2.1237  &     0.5174  &
34 &   O   &    -1.0954  &   -3.4306 &      0.9288  \\ 
3  &   Al  &    -3.9268  &    2.0726  &     1.9764  &
35 &   O   &    -1.0606  &   -0.8853 &      0.9367  \\ 
4  &   Al  &    -3.8845  &   -2.1163 &     -2.0309 &
36 &   O   &     1.3445   &   0.8895  &    -0.9797 \\ 
5  &   Al  &    -0.7291  &   -2.1526 &      2.3045  &
37 &   O   &     3.0706   &  -0.7118 &     -2.0428 \\ 
6  &   Al  &     0.9851   &   2.1587  &    -2.3391 &
38 &   O   &    -2.7743  &    0.7033  &     1.9970  \\ 
7  &   Al  &     2.1827   &  -2.0494 &      0.6751  &
39 &   O   &    -2.8279  &    3.4682  &     2.0617  \\ 
8  &   Al  &    -1.9200  &    2.1306  &    -0.6751 &
40 &   O   &     3.0547   &  -3.4826 &     -2.0462 \\ 
9  &   Al  &     3.2610   &   0.6592  &    -0.8137 &
41 &   O   &     2.8364   &   2.0514  &     3.2291  \\ 
10 &   Al  &    -2.9661  &   -0.6557 &      0.7554  &
42 &   O   &    -2.6166  &   -2.4034 &     -3.2768 \\ 
11 &   Al  &    -2.9988  &   -3.5769 &      0.7977  &
43 &   O   &    -0.8803  &   -3.5822 &      3.4350  \\ 
12 &   Al  &     3.2391   &   3.5990  &    -0.7783 &
44 &   O   &    -0.7107  &   -0.7036 &      3.4130  \\ 
13 &   Al  &     1.1980   &  -3.4664 &     -2.1325 &
45 &   O   &     1.0240   &   0.7104  &    -3.4290 \\ 
14 &   Al  &    -0.9425  &    3.4665  &     2.1387  &
46 &   O   &     1.1612   &   3.5945  &    -3.4700 \\ 
15 &   Al  &    -0.8565  &    0.6350  &     2.0730  &
47 &   O   &     1.1027   &  -2.0526 &     -3.3691 \\ 
16 &   Al  &     1.1586   &  -0.6416 &     -2.1244 & 
48 &   O   &    -0.8946  &    2.0340  &     3.3624  \\ 
17 &   O   &     1.1042   &  -2.4880 &      2.0387  & 
49 &   H   &    -0.8990  &   -3.6873 &      4.4039  \\ 
18 &   O   &    -0.8358  &    2.4934  &    -2.0475 & 
50 &   H   &    -0.1868  &   -0.5600 &      4.2270  \\ 
19 &   O   &     3.0532   &   2.1092  &     0.4986  & 
51 &   H   &    -1.6305  &    1.9482  &     4.0115  \\ 
20 &   O   &    -2.7804  &   -2.1103 &     -0.5342 & 
52 &   H   &     1.2013   &  -3.5711 &      2.0778  \\ 
21 &   O   &     1.0544   &  -2.0888 &     -0.7611 & 
53 &   H   &     3.8205   &   2.1588  &    -2.8493 \\ 
22 &   O   &    -0.7881  &    2.0931  &     0.7454  & 
54 &   H   &     1.0212   &   0.5948  &    -4.3975 \\ 
23 &   O   &     3.1987   &   2.1327  &    -2.0858 & 
55 &   H   &     0.3266   &  -2.0683 &     -3.9760 \\ 
24 &   O   &    -2.8919  &   -2.1045 &      2.0757  & 
56 &   H   &     2.9879   &   2.0437  &     4.2012  \\ 
25 &   O   &     0.9815   &   3.4223  &     1.9889  & 
57 &   H   &    -2.7926  &   -2.4397 &     -4.2435 \\ 
26 &   O   &    -0.7320  &   -3.4516 &     -1.9971 & 
58 &   H   &     1.6861   &   0.8258  &     2.5299  \\ 
27 &   O   &    -0.6912  &   -0.7763 &     -1.8274 & 
59 &   H   &    -3.4662  &   -2.0932 &      2.8751  \\ 
28 &   O   &     0.9784   &   0.7850  &     1.8281  & 
60 &   H   &    -0.9186  &    3.6123  &    -2.0851 \\ 
29 &   O   &     3.2683   &  -0.6428 &      0.5749  & 
61 &   H   &     1.7686   &   3.7268  &    -4.2273 \\ 
30 &   O   &    -2.9337  &    0.6544  &    -0.6651 & 
62 &   H   &    -1.4400  &   -0.1669 &     -2.0215 \\ 
31 &   O   &    -3.0102  &    3.5246  &    -0.5532 & 
63 &   H   &     1.6000   &   3.0011  &     2.6990  \\ 
32 &   O   &     3.2518   &  -3.4887 &      0.5769  & 
64 &   H   &    -1.4481  &   -3.0488 &     -2.6881 \\ \hline\hline
	\end{tabular}
\end{center}
\end{minipage}
\end{table}

\clearpage

\begin{figure}[!ht]
\begin{minipage}[c][1\textheight]{1\textwidth}
\centering
\includegraphics[width=0.85\hsize]{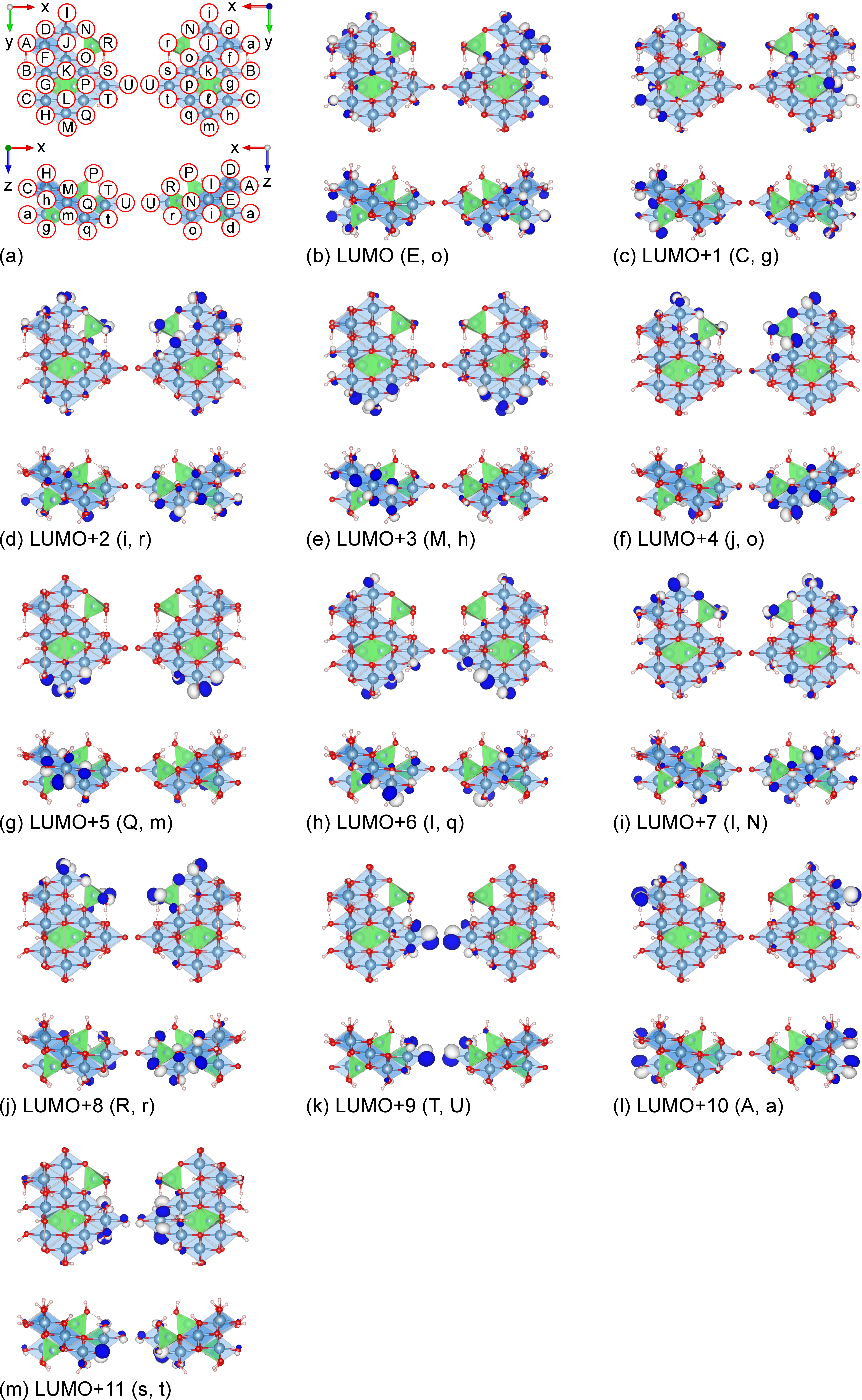}
\caption{
(a)
Labels of O atoms of the bare cluster.
(b)--(m) 
LUMO--LUMO+11 of the bare cluster.
The O atoms that should be terminated by H atoms 
because of large molecular orbital coefficients
are provided in parentheses.
Isosurface values are $5.0\times10^{-2}$. a.u.
}
  \label{FIGS2}
\end{minipage}
\end{figure}

\clearpage

\begin{figure}[!ht]
\begin{minipage}[c][1\textheight]{1\textwidth}
\centering
\includegraphics[width=0.6\hsize]{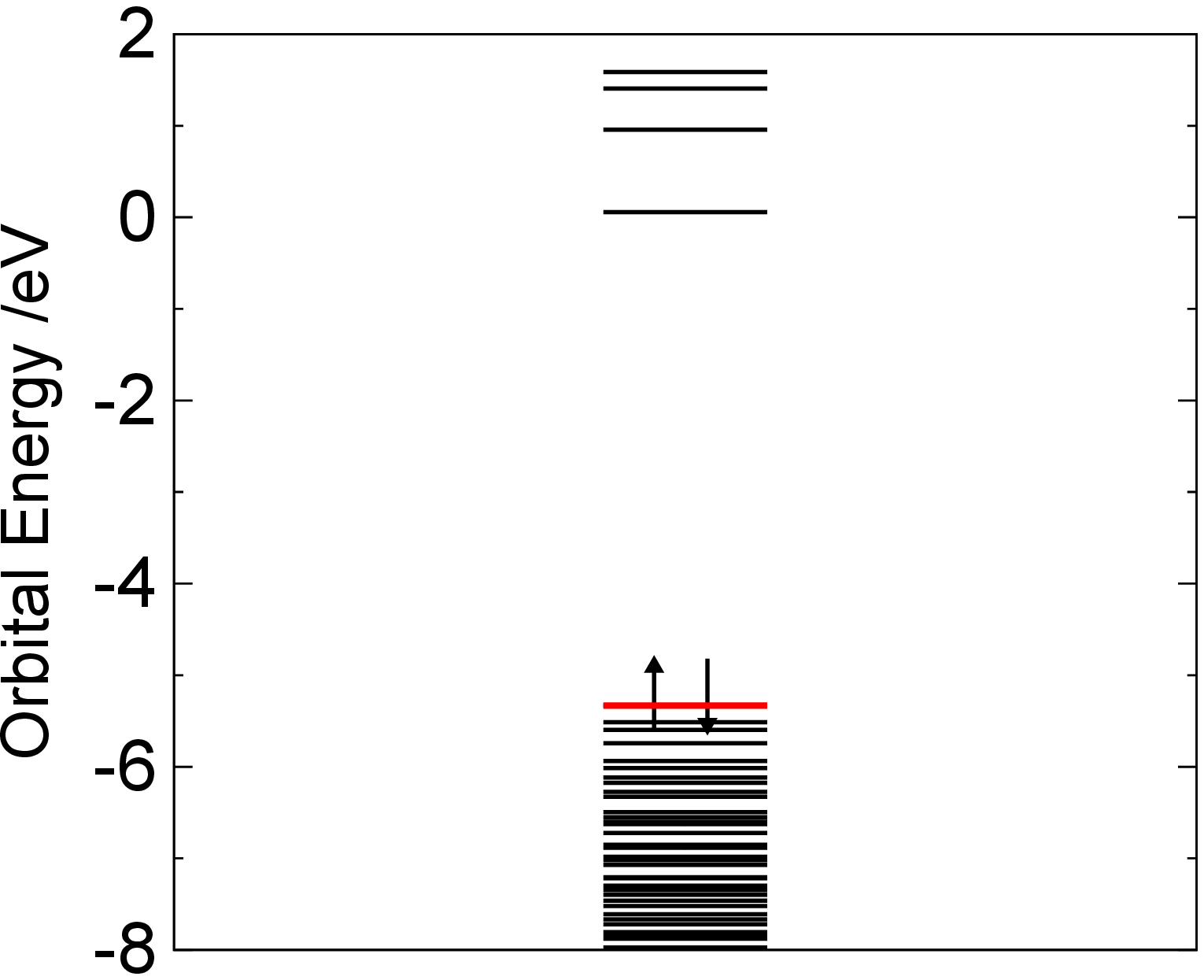}
\caption{
Calculated orbital levels of the SSHT cluster model.
The energy gap is 5.39 eV.
}
  \label{FIGS3}
\end{minipage}
\end{figure}

\clearpage

\begin{table}[!h]
\begin{minipage}[c][1\textheight]{1\textwidth}
\caption{\label{TABLES2} 
Cartesian coordinates of the SSHT cluster model.
The coordinates are provided in Angstrom units.
}
\begin{center}
	\begin{tabular}{ccrrrccrrr} \hline\hline
No. & Atom & 
\multicolumn{1}{c}{$x$} & \multicolumn{1}{c}{$y$} & \multicolumn{1}{c}{$z$} &
No. & Atom & 
\multicolumn{1}{c}{$x$} & \multicolumn{1}{c}{$y$} & \multicolumn{1}{c}{$z$} \\ \hline
 1  &  Al   & -2.5502   &  3.0942   & -0.8695 &
 48 &   O   &  5.2249   &  1.8849   &  1.2792 \\
 2  &  Al   &  3.2166   & -1.2169   &  1.6664 &
 49 &   O   & -4.6889   &  0.8051   &  2.4615 \\
 3  &  Al   & -1.5393   & -0.0745   & -1.3152 &
 50 &   O   &  5.3589   &  1.2095   & -1.4670 \\
 4  &  Al   &  3.4200   & -1.3881   & -1.2848 &
 51 &   O   & -5.1650   & -1.4838   &  1.7196 \\
 5  &  Al   & -1.0325   & -1.1988   &  1.6144 &
 52 &   O   & -4.4465   &  0.2456   & -0.1377 \\
 6  &  Al   &  1.7267   &  3.4648   & -0.4740 &
 53 &   O   & -3.2238   &  4.6504   & -1.4638 \\
 7  &  Al   &  4.1619   &  1.2338   & -0.0048 &
 54 &   H   & -0.3025   &  3.4943   & -2.2320 \\
 8  &  Al   & -3.9027   & -1.5183   &  0.3049 &
 55 &   H   &  3.3814   & -3.7740   & -1.2575 \\
 9  &  Al   & -3.3571   &  1.0729   &  1.1627 & 
 56 &   H   &  1.6778   & -1.5870   & -3.1003 \\
 10 &  Al   & -0.6281   &  1.9749   &  1.0517 &
 57 &   H   & -3.0948   &  1.4581   & -2.7084 \\
 11 &  Al   &  1.2776   &  0.2910   &  0.0178 &
 58 &   H   & -1.2750   & -5.7087   & -0.6665 \\
 12 &  Al   &  0.7433   & -2.2661   & -0.9075 &
 59 &   H   &  2.4161   & -1.2264   &  3.8830 \\
 13 &  Al   & -1.8693   & -3.2133   & -0.6837 &
 60 &   H   & -1.2305   &  2.1810   & -2.8075 \\
 14 &   O   & -0.4508   & -1.7859   &  3.2118 &
 61 &   H   &  1.7527   & -3.9919   &  0.4531 \\
 15 &   O   &  1.1962   &  4.0541   & -2.0353 &
 62 &   H   &  0.8763   & -1.6662   &  3.1542 \\
 16 &   O   &  4.0528   & -3.1077   & -0.9378 &
 63 &   H   &  4.4182   & -3.3290   &  2.1297 \\
 17 &   O   &  2.7527   &  0.7012   &  1.2228 &
 64 &   H   &  2.6147   &  0.9433   & -2.0955 \\
 18 &   O   & -1.9346   &  1.3556   & -0.2431 &
 65 &   H   & -0.0233   &  3.1316   &  3.2074 \\
 19 &   O   & -0.5662   & -2.3646   &  0.3723 &
 66 &   H   &  0.1691   & -4.6243   & -0.3052 \\
 20 &   O   &  2.8207   &  0.5138   & -1.2330 &
 67 &   H   & -0.7992   & -2.6022   &  3.6363 \\
 21 &   O   & -5.0136   & -2.4729   & -0.9316 &
 68 &   H   & -5.3616   & -1.8834   & -1.6464 \\
 22 &   O   & -2.0283   &  1.8322   &  2.3045 &
 69 &   H   &  1.7379   &  5.5453   &  0.9977 \\
 23 &   O   &  1.9749   & -1.4832   &  3.0373 &
 70 &   H   &  5.2130   & -1.4785   & -3.0162 \\
 24 &   O   & -1.2246   &  3.0410   & -2.2778 &
 71 &   H   &  6.1809   &  1.6539   &  1.3343 \\
 25 &   O   &  1.3252   & -4.0322   & -0.4446 &
 72 &   H   & -4.2572   & -3.2741   & -1.4077 \\
 26 &   O   & -3.7172   &  2.0835   & -2.1075 &
 73 &   H   & -5.4722   &  1.4035   &  2.3806 \\
 27 &   O   & -0.9863   & -4.8499   & -0.2816 &
 74 &   H   & -3.3846   & -3.8344   &  1.3581 \\
 28 &   O   &  4.4445   & -0.7267   &  2.8799 &
 75 &   H   & -0.3270   & -3.6149   & -2.7614 \\
 29 &   O   & -2.6935   & -0.6290   &  1.4533 &
 76 &   H   &  5.1581   &  0.2420   & -2.1452 \\
 30 &   O   &  3.4820   &  2.9984   & -0.3267 &
 77 &   H   &  2.4791   &  1.2776   &  1.9776 \\
 31 &   O   &  0.8436   &  2.0471   & -0.0863 &
 78 &   H   & -1.8570   &  0.0726   & -3.7532 \\
 32 &   O   & -0.0033   &  0.2405   &  1.3385 &
 79 &   H   &  4.0737   &  3.5486   &  0.2505 \\
 33 &   O   &  2.1207   & -1.4103   &  0.1288 &
 80 &   H   &  4.8517   &  0.1673   &  2.7492 \\
 34 &   O   & -1.2723   &  3.6264   &  0.3936 &
 81 &   H   & -4.5038   &  0.7685   & -0.9782 \\
 35 &   O   & -3.8041   &  2.8138   &  0.5731 &
 82 &   H   & -5.0325   & -0.4170   &  2.2211 \\
 36 &   O   &  4.4883   & -0.6683   &  0.1399 &
 83 &   H   & -3.0618   & -3.6900   & -2.7464 \\
 37 &   O   & -3.1081   & -3.1683   &  0.7013 &
 84 &   H   &  0.9085   &  3.7835   &  1.9153 \\
 38 &   O   & -2.5618   & -1.4750   & -1.0319 &
 85 &   H   &  5.5914   &  2.0413   & -1.9385 \\
 39 &   O   &  0.1992   & -0.4509   & -1.2645 &
 86 &   H   &  1.4333   &  4.8380   & -2.5729 \\
 40 &   O   & -1.9214   &  0.6541   & -2.9605 &
 87 &   H   & -1.9424   &  1.2946   &  3.1263 \\
 41 &   O   &  4.6498   & -0.8188   & -2.5521 &
 88 &   H   & -3.9527   &  5.0606   & -0.9370 \\
 42 &   O   & -3.2263   & -3.9387   & -1.8010 &
 89 &   H   & -4.2246   &  2.7701   & -2.6092 \\
 43 &   O   &  2.0495   & -2.0500   & -2.3176 &
 90 &   H   &  5.4101   & -0.9795   &  0.2964 \\
 44 &   O   & -0.5691   & -2.9512   & -2.0781 &
 91 &   H   &  4.0413   & -3.1975   &  0.2902 \\
 45 &   O   &  3.7946   & -2.9858   &  1.4495 &
 92 &   H   & -3.6800   &  3.4878   &  1.2846 \\
 46 &   O   &  0.4584   &  2.9261   &  2.3711 &
 93 &   H   & -0.8927   &  4.5192   &  0.5244 \\
 47 &   O   &  1.4592   &  4.6146   &  0.8687 &
 94 &   H   & -6.1008   & -1.6834   &  1.4785 \\ 
 \hline \hline
      \end{tabular}                                        
\end{center}                                              
\end{minipage}
\end{table}

\clearpage 

~\\
\vfill

\begin{table}[!ht]
\vspace*{-\intextsep}
\caption{\label{TABLES3} 
OVCCs for the effective mode of the $\gamma$-alumina cluster model.
The core orbitals are neglected.
}
\begin{center}
\begin{tabular}{cccc} \hline
	$i$            &          &   Energy / eV & $f_{i,\xi} \times10^{-3}$ a.u. \\ \hline
	265            & HOMO     &   -5.921      & 1.525  \\
	263            & HOMO-2   &   -8.861      & 1.052  \\
	264            & HOMO-1   &   -8.978      & 1.038  \\
	260            & HOMO-5   &   -9.161      & 0.840  \\
	\vdots         & \vdots   &   \vdots      & \vdots  \\
	149            & HOMO-116 &  -17.613      &  0.007  \\
\hline
\end{tabular}
\end{center}
\vspace*{-\intextsep}
\end{table}

\vfill

\clearpage

\begin{figure}[!ht]
\vspace*{-\intextsep}
\begin{minipage}[c][1\textheight]{1\textwidth}
\centering
\includegraphics[width=0.5\hsize]{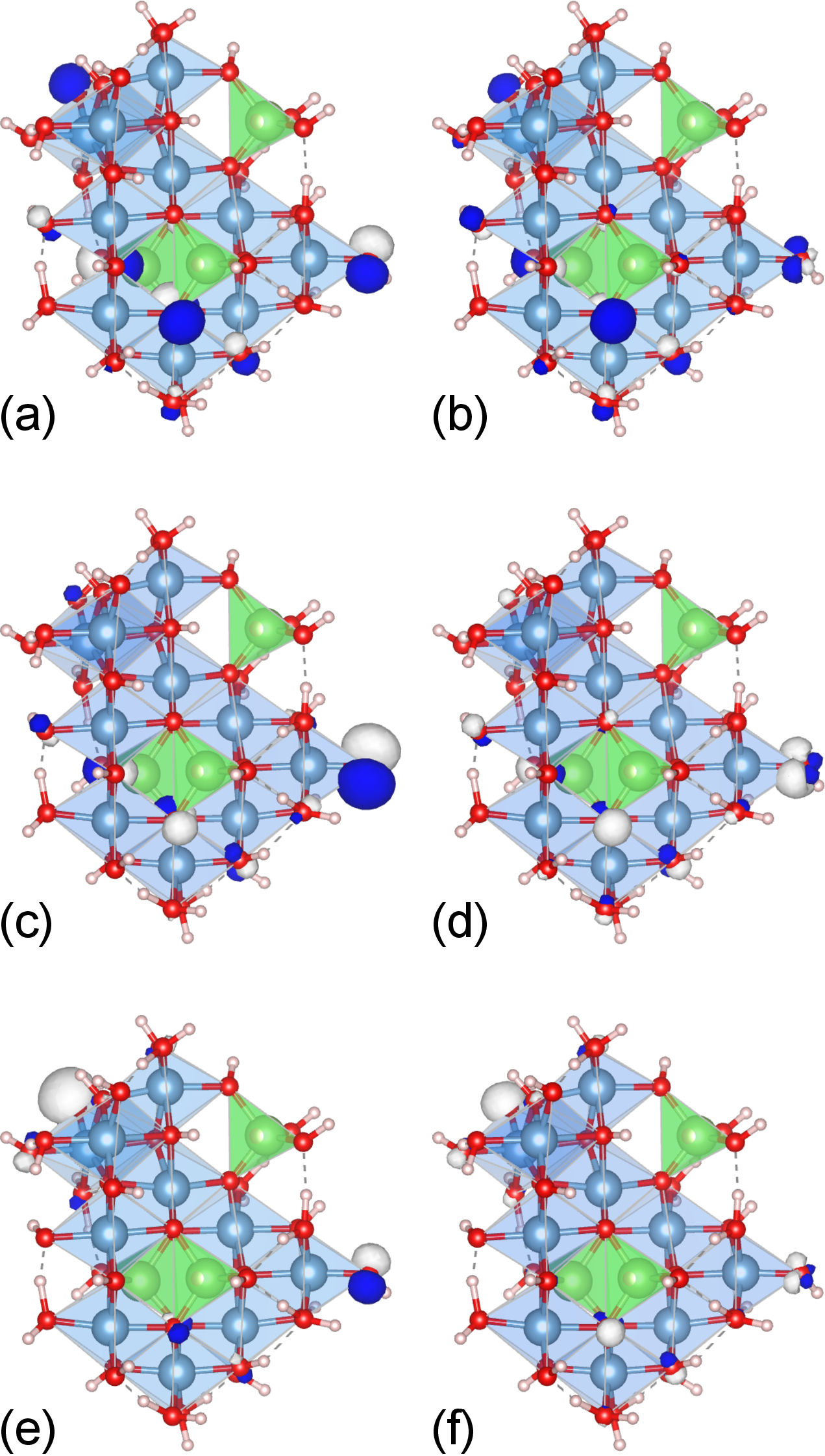}
\caption{
(a) HOMO,
(b) OVCD for the HOMO, 
(c) HOMO-2,
(d) OVCD for the HOMO-2,
(e) HOMO-1, and
(f) OVCD for the HOMO-1
of the $\gamma$-alumina cluster model.
Isosurface values of the molecular orbitals are $5.0\times10^{-2}$ a.u.,
and of the OVCDs are $2.0\times10^{-5}$ a.u.
}
\label{FIGS4}
\end{minipage}
\vspace*{-\intextsep}
\end{figure}

\vfill

\begin{figure}[!ht]
\begin{minipage}[c][1\textheight]{1\textwidth}
\centering
\includegraphics[width=0.6\hsize]{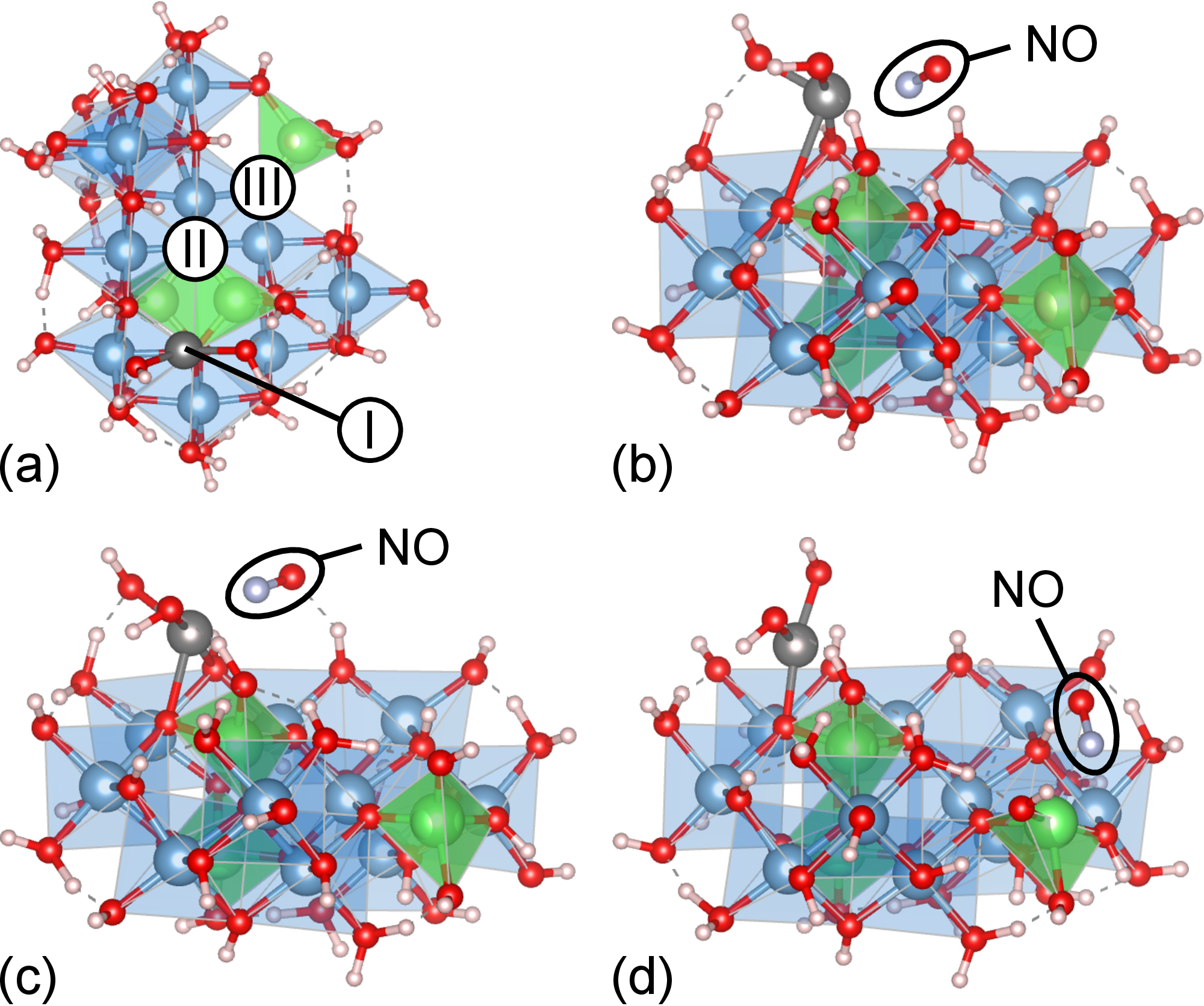}
\caption{
(a) 
Initial positions of NO on the Cu(OH)$_2$/$\gamma$-alumina cluster model.
Geometry-optimized structures obtained by initially placing NO on positions
(b) I, (c), II, or (d) III.
The total energies of (c) and (d) respectively are 0.261 and 1.416 eV
higher than that of (b).
The reference of the total energy is that of (b), 
-8068.23720195 a.u.
}
\label{FIGS5}
\end{minipage}
\end{figure}

\vfill

\clearpage

\begin{figure}[!ht]
\begin{minipage}[c][1\textheight]{1\textwidth}
\centering
\includegraphics[width=0.6\hsize]{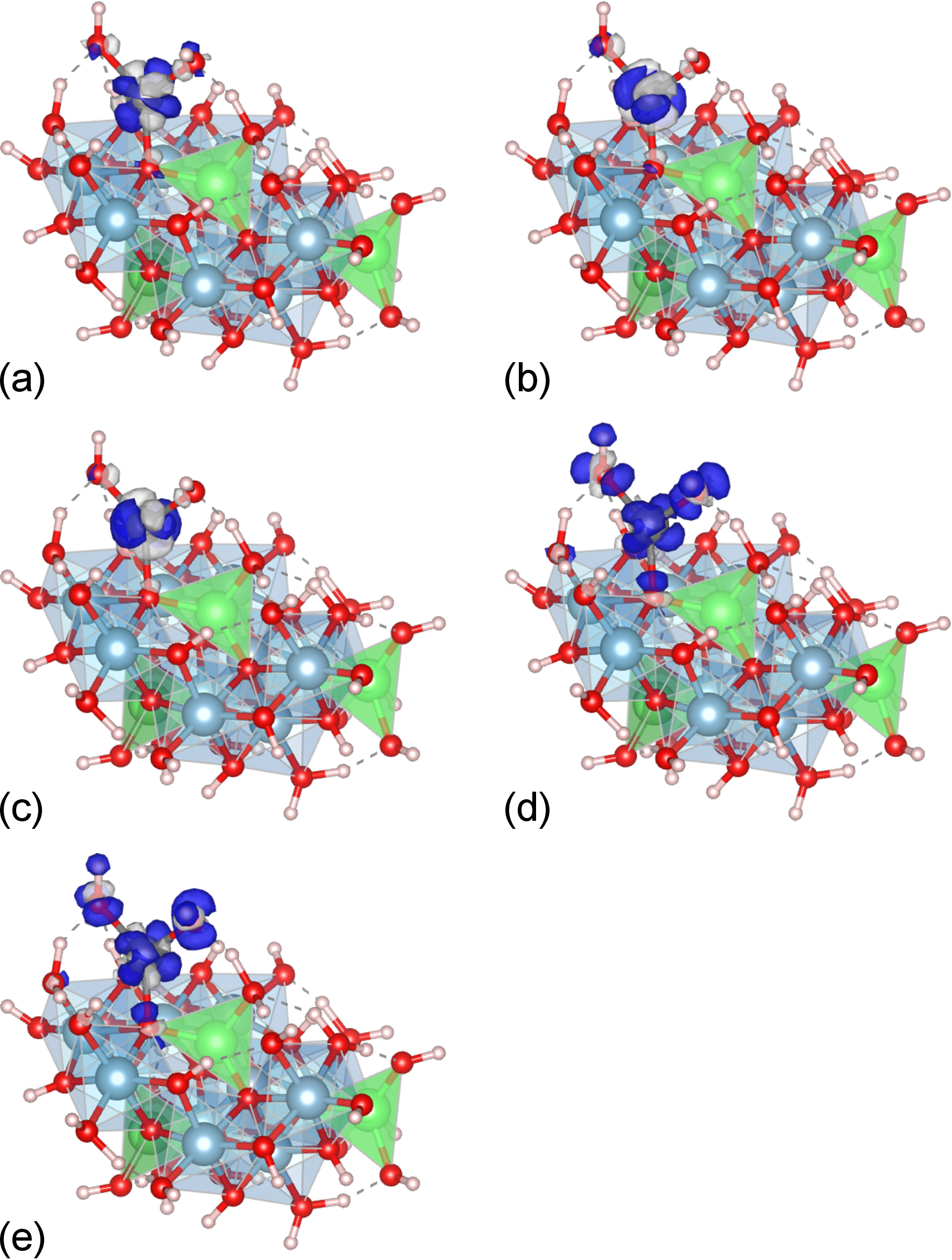}
\caption{
Electron density differences of the Cu(OH)$_2$/$\gamma$-alumina cluster model 
between
(a) S$_1$ and S$_0$,
(b) S$_2$ and S$_0$,
(c) S$_4$ and S$_0$,
(d) S$_6$ and S$_0$, and
(e) S$_{12}$ and S$_{0}$.
Isosurface values are $5.0\times10^{-3}$ a.u.
}
\label{FIGS6}
\end{minipage}
\end{figure}

\clearpage

\begin{figure}[!ht]
\begin{minipage}[c][1\textheight]{1\textwidth}
\centering
\includegraphics[width=0.6\hsize]{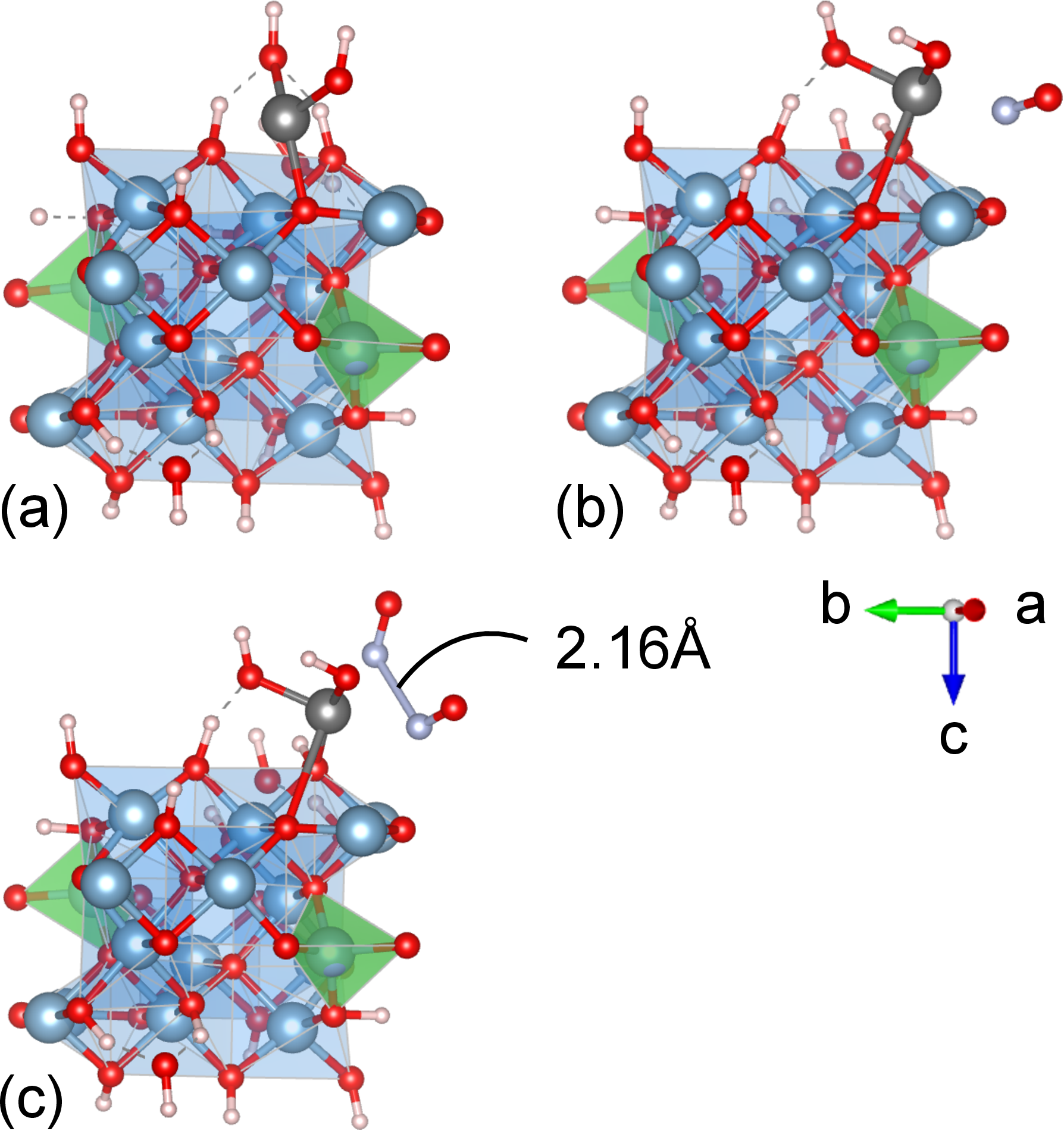}
\caption{
Geometry-optimized structures of the
(a) Cu(OH)$_2$/$\gamma$-alumina,
(b) NO-adsorbed Cu(OH)$_2$/$\gamma$-alumina, and
(c) (NO)$_2$-adsorbed Cu(OH)$_2$/$\gamma$-alumina
slab models.
}
\label{FIGS7}
\end{minipage}
\end{figure}

\clearpage

~\\

\vfill
\begin{figure}[!ht]
\vspace*{-\intextsep}
\centering
\includegraphics[width=0.6\hsize]{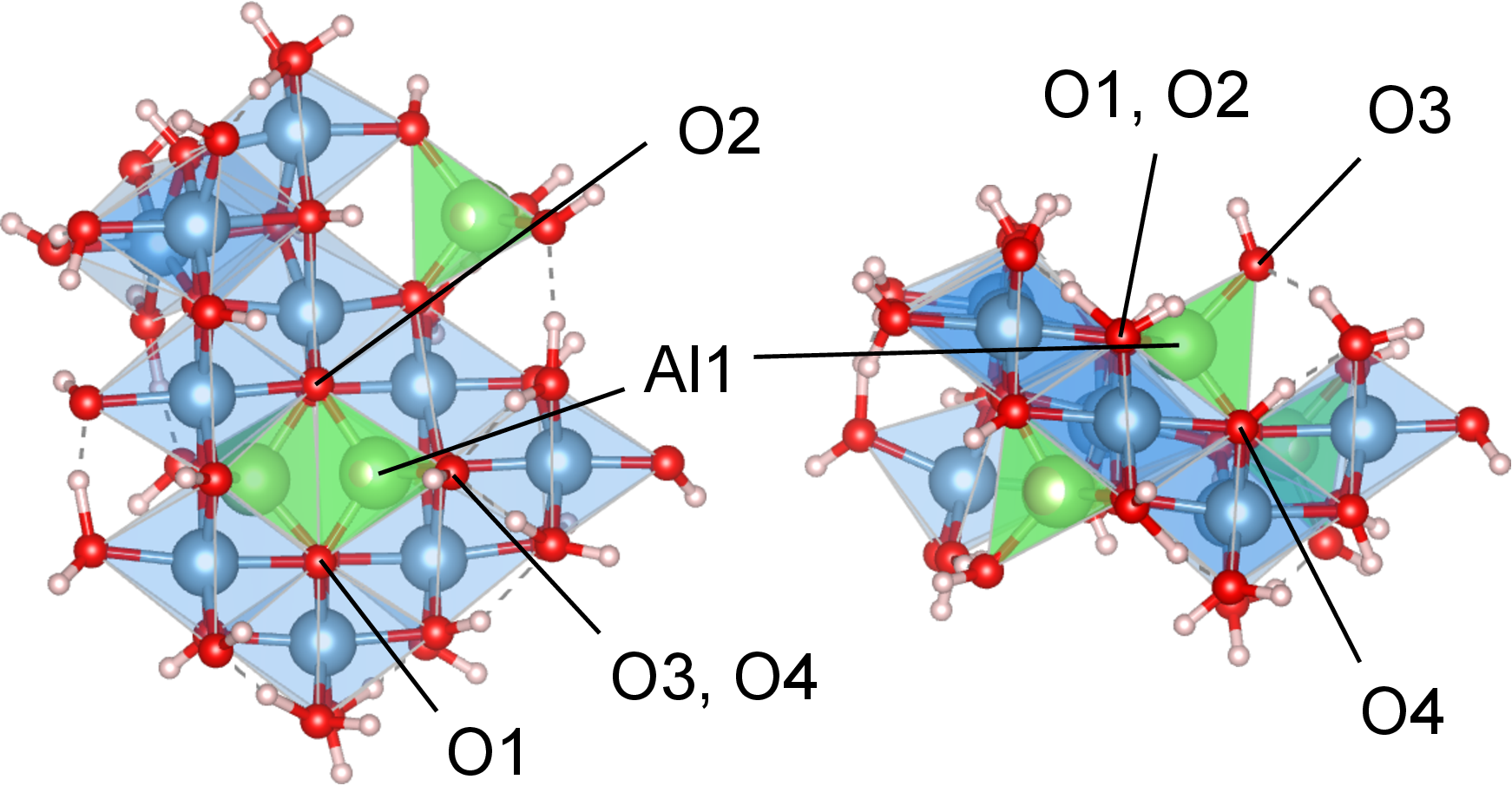}
\caption{
Atomic labels of the tetrahedral Al species
on which Cu(OH)$_2$ is adsorbed.
}
\label{FIGS8}
\vspace*{-\intextsep}
\end{figure}

\vfill

\begin{table}[!h]
\vspace*{-\intextsep}
\caption{\label{TABLES4} 
Interatomic distance and Mulliken charge of the tetrahedral Al species
in Fig.~S8 before and after the Cu adsorption.
}
\centering
\scalebox{0.9}{
\begin{tabular}{ccccccccccc} 
\hline\hline
              &        & \multicolumn{4}{c}{$d$ (\AA)}           & \multicolumn{5}{c}{Mulliken charge} \\
              &        & Al1-O1 & Al1-O2 &Al1-O3 & Al1-O4  & Al1 & O1 & O2 & O3 & O4 \\
\hline                                    
Cluster model & $\gamma$-alumina       & 1.757  & 1.779  &1.839  & 1.830  & 0.928 & -0.876 & -0.860 & -0.723 & -0.920  \\
              & Cu/$\gamma$-alumina    & 1.823  & 1.775  &1.819  & 1.804  & 1.043 & -0.877 & -0.864 & -0.799 & -0.935  \\
Slab model    & $\gamma$-alumina       & 1.791  & 1.772  &1.800  & 1.859  & 1.311 & -1.004 & -1.017 & -0.931 & -1.008  \\
              & Cu/$\gamma$-alumina    & 1.845  & 1.753  &1.785  & 1.837  & 1.436 & -0.898 & -1.026 & -0.948 & -1.018  \\
\hline\hline
\end{tabular}
}
\vspace*{-\intextsep}
\end{table}

\vfill

\clearpage

~\\

\vfill
\begin{figure}[!ht]
\centering
\includegraphics[width=0.5\hsize]{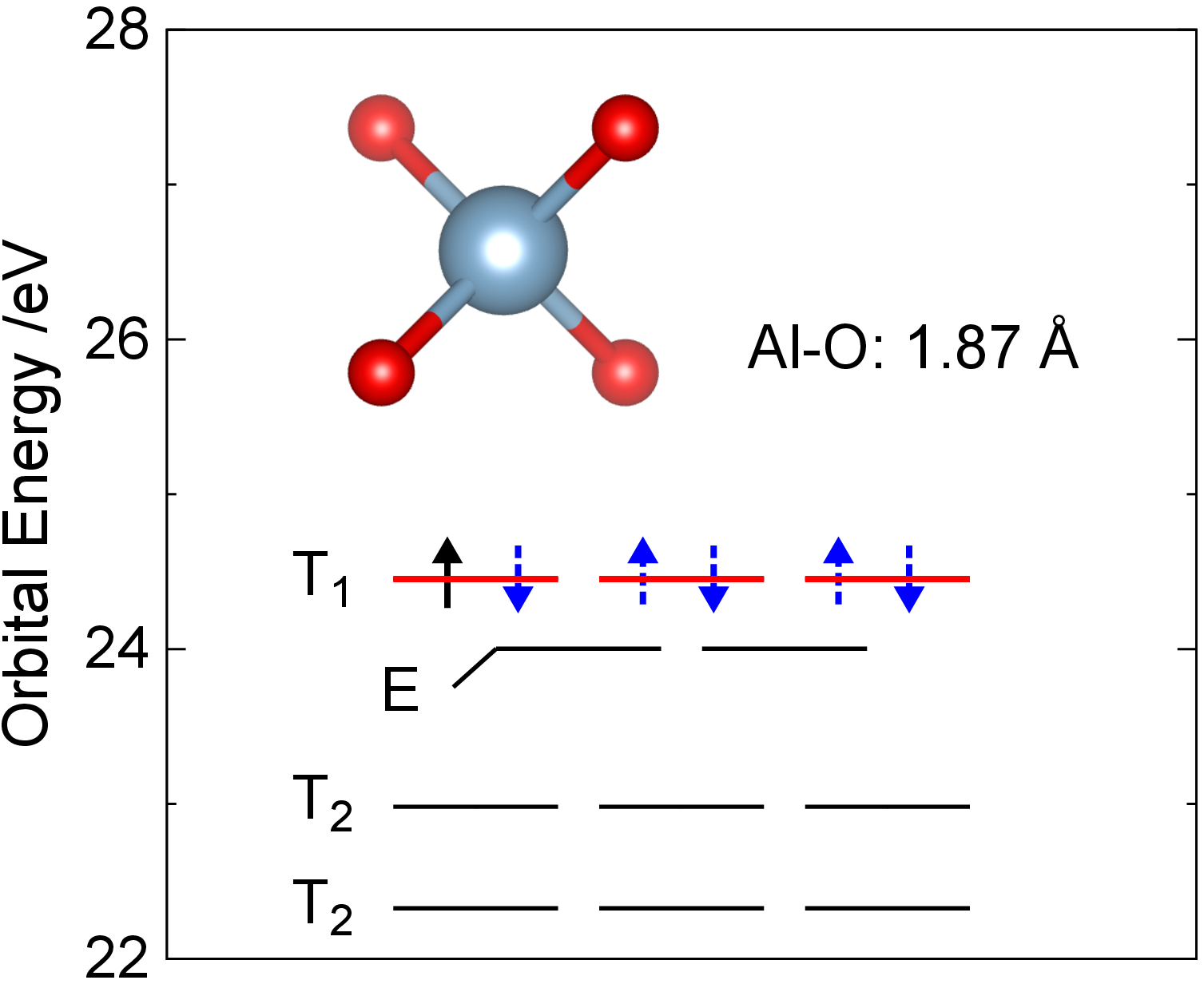}
\caption{
Orbital levels of the geometry-optimized AlO$_4$
with the $T_d$ symmetry.
Five electrons are provided to obtain the appropriate oxidation number.
}
\label{FIGS9}
\end{figure}

\vfill

\begin{table}[!h]
\caption{\label{TABLES5}
VCCs of AlO$_4$.
}
\centering
\begin{tabular}{cccccccc}
\hline\hline
      & Frequency (cm$^{-1}$) & VCC (10$^{-4}$ a.u.) \\
\hline
$e$   & 267.1                 & 0.0000                \\
$t_2$ & 336.0                 & 0.8621                \\
$a_1$ & 509.0                 & 2.1265                \\
$t_2$ & 582.9                 & 1.1002                \\
\hline\hline
\end{tabular}
\end{table}

\vfill

\clearpage

~\\
\vfill
\begin{figure}[!ht]
\centering
\includegraphics[width=0.7\hsize]{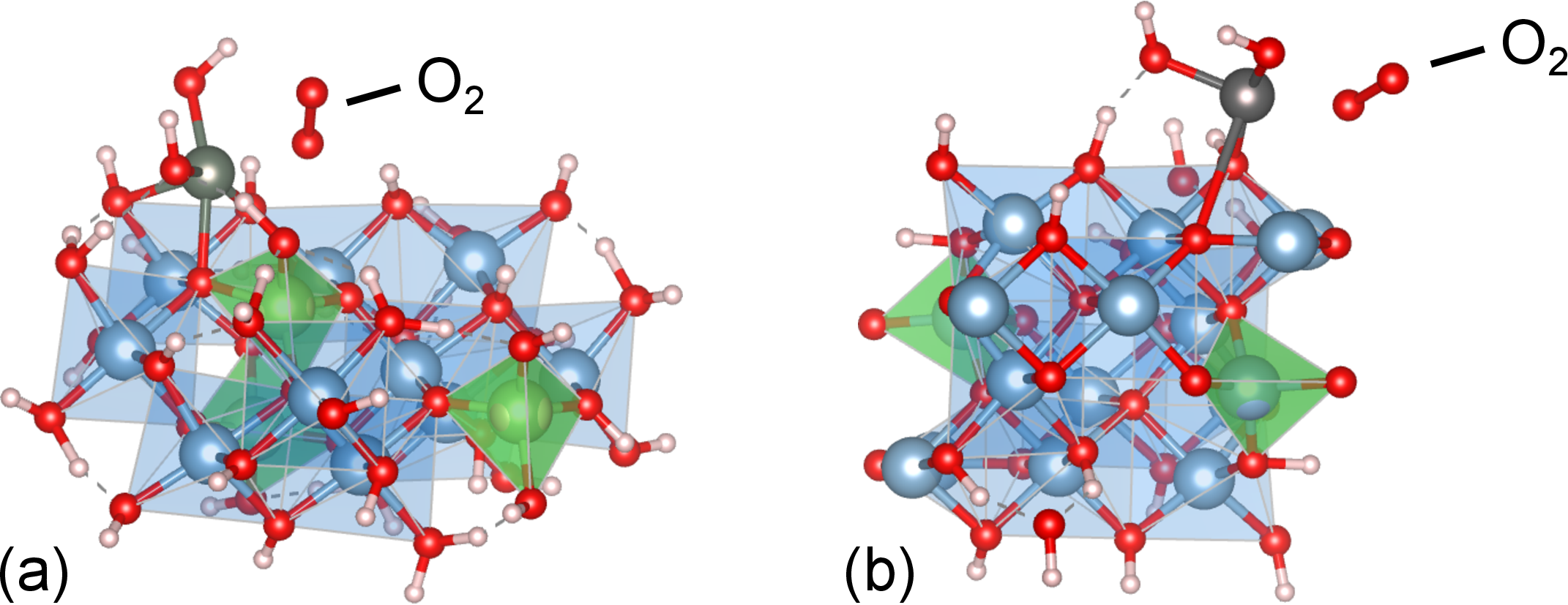}
\caption{
Geometry-optimized structures of the O$_2$-adsorbed Cu(OH)$_2$/$\gamma$-alumina 
(a) cluster and (b) slab models.
}
\label{FIGS10}
\end{figure}

\vfill

\begin{figure}[!ht]
\centering
\includegraphics[width=0.8\hsize]{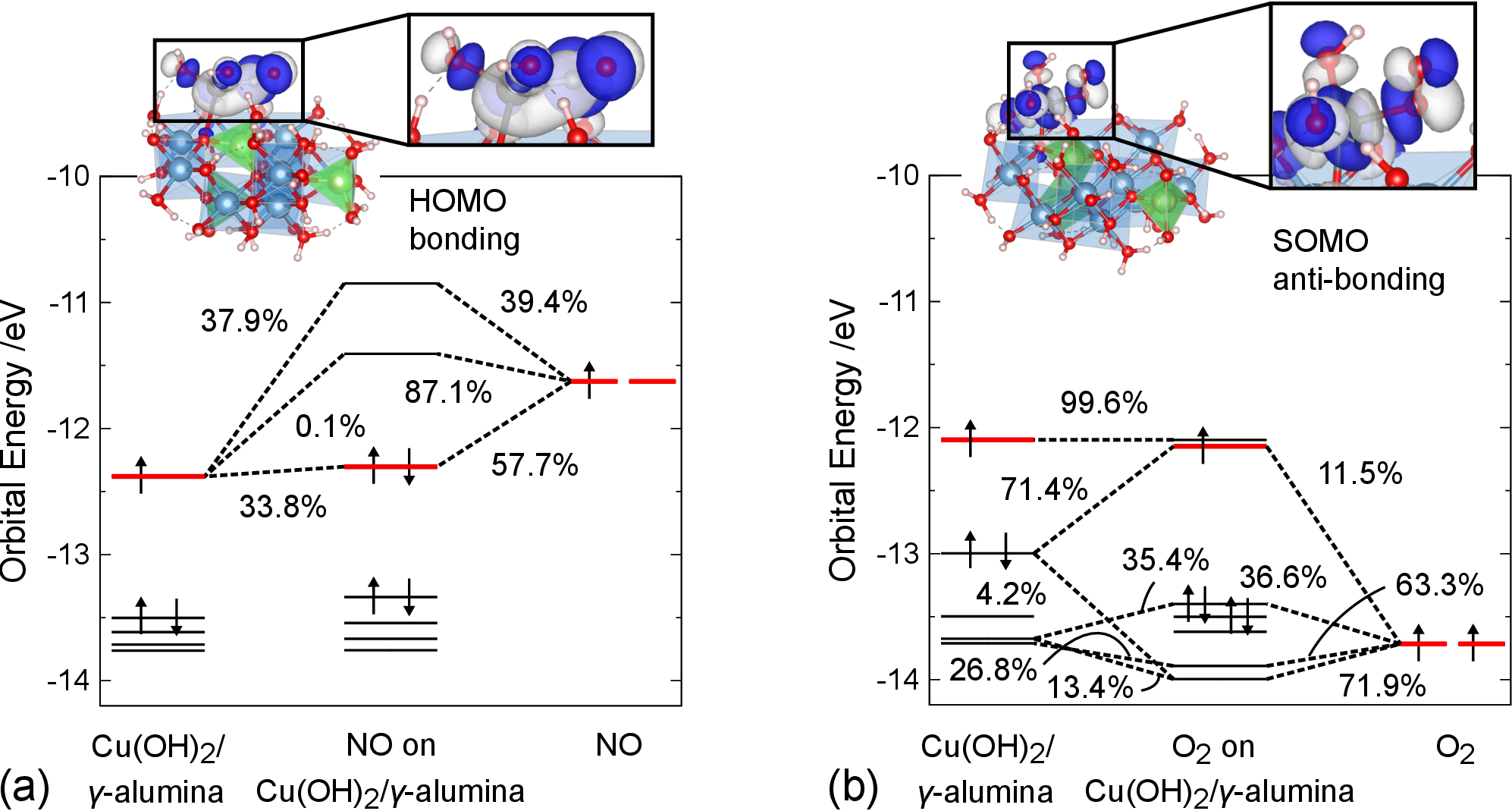}
\caption{
Fragment orbital analyses of the 
(a) NO- and (b) O$_2$-adsorbed Cu(OH)$_2$/$\gamma$-alumina cluster models. 
Values of $P_{km}$ are provided in percentages. 
}
\label{FIGS11}
\end{figure}

\vfill